\documentclass[prl,aps,showpacs,twocolumn,floatfix,superscriptaddress]
	{revtex4-1}
\usepackage{amsmath, amsfonts, amssymb, bm, graphicx}

\newcommand{\e}{{\rm e}}
\newcommand{\ii}{{\rm i}}

\newcommand{\ed}{\hat{e}^\dag}

\newcommand{\fd}{\hat{f}^\dag}
\newcommand{\gd}{\hat{g}^\dag}

\newcommand{\vac}{\vert {\rm vac} \rangle}

\begin{document}

\title{Topological superfluidity with repulsive alkaline-earth atoms in optical
lattices}

\author{L. Isaev}
\author{A. Kaufman}
\affiliation{
 JILA, NIST, Department of Physics \& Center for Theory of
 Quantum Matter, University of Colorado, 440 UCB, Boulder, CO 80309, USA
}
\author{G. Ortiz}
\email{Corresponding author, email: ortizg@indiana.edu}
\affiliation{
 Department of Physics and Center for Exploration of Energy and Matter,
 Indiana University, Bloomington IN 47405, USA
}
\author{A. M. Rey}
\email{Corresponding author, email: arey@jilau1.colorado.edu}
\affiliation{
 JILA, NIST, Department of Physics \& Center for Theory of
 Quantum Matter, University of Colorado, 440 UCB, Boulder, CO 80309, USA
}

\begin{abstract}

 Topological superfluids are of technological relevance since they are believed
 to host Majorana bound states, a powerful resource for quantum computation and
 memory.
 Here we propose to realize topological superfluidity with fermionic atoms in
 an optical lattice.
 We consider a situation where atoms in two internal states experience
 different lattice potentials: one species is localized and the other
 itinerant, and show how quantum fluctuations of the localized fermions give
 rise to an attraction and strong spin-orbit coupling in the itinerant band.
 At low temperature, these effects stabilize a topological superfluid of mobile
 atoms even if their bare interactions are repulsive.
 This emergent state can be engineered with ${}^{87}$Sr atoms in a superlattice
 with a dimerized unit cell.
 To probe its unique properties we describe protocols that use high spectral
 resolution and controllability of the Sr clock transition, such as
 momentum-resolved spectroscopy and supercurrent response to a synthetic
 (laser-induced) magnetic field.

\end{abstract}

\maketitle

\paragraph*{Introduction.--}

Our understanding of many-body systems traditionally relies on the Landau
classification of ordered states of matter based on global symmetries
spontaneously broken within a given phase.
This symmetry breaking is accompanied by emergence of an order-parameter (OP),
i.e. non-zero expectation value of a local physical observable that uniquely
characterizes the phase.
For instance a hallmark signature of a fermionic superfluid (SF) is breaking of
the particle number conservation [$U (1)$] symmetry which occurs as a result of
Cooper pairing.
The corresponding OP plays the role of a Cooper pair wavefunction and defines
an energy gap in the excitation spectrum, allowing dissipationless particle
currents \cite{landau-vol-9}.
However, many phases of matter defy the Landau paradigm.
An important class of such systems are topological superfluids (TSFs)
\cite{qi-2011-1,nayak-2008-1}, i.e. phases that in addition to $U (1)$, break a
residual $\mathbb{Z}_2$ symmetry.
The latter symmetry breaking is a global phenomenon that occurs {\it in the
absence of local OP} and only for appropriate boundary conditions
\cite{ortiz-2014-1,wen-2015-1,nussinov-2009-1}.

Despite all efforts dedicated to the search for TSFs, they remain elusive with
the only confirmed realization being liquid ${}^3 {\rm He}$-A
\cite{autti-2016-1}.
One reason for such scarcity is that TSFs require a very particular orbital
structure of Cooper pairs, at least $p$-wave, which originates either from
strongly spin-dependent interactions or a large spin-orbit coupling (SOC) that
couples particle's motion to its spin.
Apparently, the coexistence of a sizeable SOC and attractive interactions
(leading to Cooper pairing) is quite rare in nature \cite{smidman-2017-1},
fundamentally because SOC and fermion pairing have very different physical
origins.

In the present work, we propose a pathway towards topological superfluidity,
which completely overcomes the above limitation by using {\it the same}
ingredients to engineer attractive interactions and SOC.
We study a model of repulsive fermions in two bands: one localized and another
itinerant, and show that inhomogeneities spanning few lattice sites (e.g.
dimerization) in the localized band lead to two profound phenomena.
First, they induce {\it short-range} attractive interactions among the
itinerant species, by virtue of local quantum fluctuations.
Second, they enlarge the unit cell in accordance with the extent of localized
wavefunctions.
As a result, itinerant states reside in several Bloch bands, whose index plays
the role of a spin degree of freedom.
These pseudospins flip whenever a fermion tunnels between unit cells, thus
coupling to the orbital motion.
The strength of this emergent SOC is determined by the tunneling amplitude and
is {\it of the order of the itinerant bandwidth}.
We show that a combination of this {\it ultra-strong} SOC and attractive
interactions gives rise to a robust $p$-wave TSF in quasi-one dimension (1D)
and a chiral $p_x + {\rm i} p_y$ SF in 2D.

Our TSF state can be observed in ultracold {\it nuclear-spin polarized}
fermionic alkaline-earth atoms (AEAs) \cite{cazalilla-2014-1}, e.g.
${}^{87}{\rm Sr}$ \cite{zhang-2014-1} or ${}^{173}{\rm Yb}$
\cite{cappellini-2014-1,scazza-2014-1}, in an optical superlattice with a
few-site unit cell
\cite{sebby-2006-1,lee-2007-1,folling-2007-1,trotzky-2008-1}.
The localized (itinerant) states can be implemented with atoms in an excited
${}^3 P_0$ (ground state ${}^1 S_0$) clock state (respectively, $e$- and
$g$-states), with a {\it single $e$-atom per unit cell}.
We propose several experimental probes for characterizing the TSFs, including
momentum-resolved spectroscopy \cite{stewart-2008-1,dao-2009-1} and generation
of a particle supercurrent with a laser-induced synthetic magnetic field
\cite{gorkov-2001-1,yip-2002-1,ojanen-2012-1}.
Our approach avoids many known experimental issues:
(i) the only relevant interactions occur through the $a_{e g}^-$ channel, and
therefore the system is not affected by inelastic $e$-$e$ losses
\cite{bishof-2011-1,lemke-2011-1} or strong scattering in the $a_{e g}^+$
channel \cite{hofer-2015-1,pagano-2015-1};
(ii) $p$-wave interactions in our case emerge as a result of quantum
fluctuations as opposed to a $p$-wave Feshbach resonance, and our setup is free
from the three-body losses reported in experiments
\cite{regal-2003-1,zhang-2004-1,gaebler-2007-1,chin-2010-1};
(iii) the SOC in our system is generated as a result of the lattice structure
and hence avoids heating, inherent to earlier proposals to create SOC using
near-resonant Raman lasers
\cite{galitski-2013-1,goldman-2014-1,celi-2014-1,dalibard-2011-1}.
Our proposal is much simpler than previous works that involve more complicated
laser arrays, RF pulses or additional molecular states
\cite{cooper-2009-1,kraus-2013-1,buhler-2014-1,liu-2014-1,wang-2016-1,
iemini-2017-1}.
Finally, our cold-atom system provides a long-sought-after realization of
a pairing mechanism in repulsive fermions, which emerges because of nanoscale
inhomogeneities \cite{eroles-2000-1,tsai-2008-1,rey-2009-1,isaev-2010-1}.

\begin{figure}[t]
 \begin{center}
  \includegraphics[width = \columnwidth]{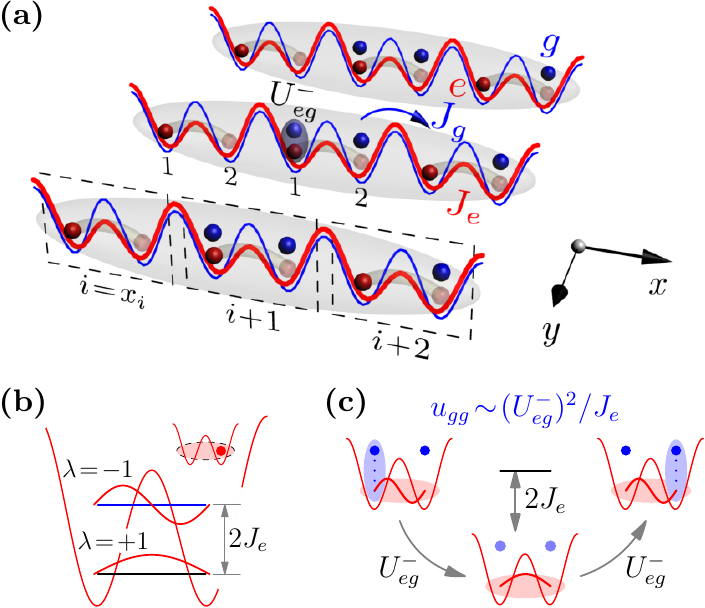}
 \end{center}
 \caption{
  {\bf (a)} The system described by Eq. \eqref{eq:H_1d} can be implemented by
  tightly confining in an array of 1D tubes an ultra-cold gas of nuclear-spin
  polarized fermionic alkaline-earth atoms prepared in the clock states $g$ (in
  blue) and $e$ (red color).
  Along the tubes, $e$-atoms experience a superlattice that consists of
  weakly-coupled double-wells (dimers) with large intra-dimer tunneling $J_e$.
  The $g$-atoms are itinerant and experience a weaker lattice potential along
  the tube direction with a nearest-neighbor hopping $J_g$.
  Within each tube, a unit cell (dashed rectangle) at a position $x_i = i$
  includes two lattice sites labeled with $a = 1$, $2$ (4 wells overall).
  There is a $e$-$g$ repulsive interaction $U^-_{e g} > 0$, assumed small
  compared to $J_e$: $U^-_{e g} \ll J_e$.
  {\bf (b)} Symmetric ($\lambda = +$) and antisymmetric ($\lambda = -$)
  $e$-atom kinetic-energy eigenstates within a dimer.
  {\bf (c)} When $e$-atom dimers are prepared in the anti-symmetric mode,
  virtual transitions to the symmetric state, caused by the $e$-$g$
  interaction, induce an effective attraction $u_{g g} = (U^-_{e g})^2 / 4 J_e$
  between two $g$-fermions within a dimer.
  These processes are captured by the effective model \eqref{eq:H_ef-1d}.
 }
 \label{fig:fig1}
\end{figure}

\paragraph*{$p$-wave SF in a 1D superlattice.--}

Key aspects of the emergent Cooper pairing and SOC leading to our proposed TSF
state can be seen by studying a dimerized 1D optical lattice, shown in Fig.
\ref{fig:fig1}(a) and described by the model Hamiltonian:
\begin{align}
 \hat{H} = & -J_e \sideset{}{_i} \sum \bigl( \ed_{i 1} \hat{e}_{i 2} +
 {\rm h.c.} \bigr) + U^-_{e g} \sideset{}{_{i a}} \sum \hat{n}^e_{i a}
 \hat{n}^g_{i a} \nonumber \\
 & - J_g \sideset{}{_i} \sum \bigl( \gd_{i 1} \hat{g}_{i 2} + \gd_{i + 1, 1}
 \hat{g}_{i 2} + {\rm h. c.} \bigr),
 \label{eq:H_1d}
\end{align}
where $i = x_i = 0, \ldots, N_d - 1$ and $a = 1, 2$ labels dimers and sites
within a dimer, respectively.
The operator $\ed_{i a}$ ($\gd_{i a}$) creates a {\it nuclear-spin polarized}
$e$ ($g$) atom at site $a$ within a dimer $i$ ($\hat{n}^e_{i a} = \ed_{i a}
\hat{e}_{i a}$ and similarly for $\hat{n}^g_{i a}$).
The $e$-atoms occupy a dimerized lattice with a large intra-dimer hopping
$J_e > 0$ and one atom per dimer (we assume that dimers are decoupled).
The $g$-atoms propagate in a simple (non-dimerized) lattice with a
nearest-neighbor tunneling $J_g$.
The second term in \eqref{eq:H_1d} contains a local $e$-$g$ repulsion of
strength $U^-_{e g} \! > \! 0$.

We focus on the regime $J_e \gg U^-_{e g}$ and $J_g$ when interactions and
$g$-atom kinetic energy in \eqref{eq:H_1d} can be considered a perturbation to
the $e$-atom kinetic energy.
For $i$-th dimer, the latter has eigenstates $\vert \lambda \rangle_i =
\frac{1}{\sqrt{2}} (\ed_{i 1} + \lambda \ed_{i 2}) \vac$ ($\lambda = \pm 1$ and
$\vac$ is the vacuum state without atoms) with energies $-\lambda J_e$ [Fig.
\ref{fig:fig1}(b)].
States of the entire $e$-$g$ system can be approximately written as
$\vert \Psi_{e g} \rangle = \prod_i \vert \lambda \rangle_i \otimes \vert
\Psi_g \rangle$ ($\vert \Psi_g \rangle$ is a state of only $g$-atoms), thanks
to the single-dimer gap $2 J_e$.

We next assume that $e$-subsystem is prepared in the excited state $\prod_i
\vert \lambda = -1 \rangle_i$.
This configuration is stable because of the large energy penalty $2 J_e$ that
suppresses decay of individual dimers to their ground state (GS) with $\lambda
= +1$ in the absence of decoherence sources (this requirement is well satisfied
in cold-atom systems), for instance due to $e$-$g$ scattering.
The weak interactions $U^-_{e g}$ only induce $e$-atom virtual transitions to
dimer states with $\lambda = +1$, which we take into account via 2nd order
perturbation theory (the kinetic energy of $g$-atoms amounts to a 1st order
correction because it operates within the degenerate subspace $\lbrace \vert
\Psi_{e g} \rangle \rbrace$).
These virtual processes, shown in Fig. \ref{fig:fig1}(c), give rise to an
effective Hamiltonian for the $g$-subsystem (see Methods)
\begin{align}
 \hat{H}_{\rm ef} = & \hat{H}^g_0 - u_{g g} \sideset{}{_i} \sum
 \hat{n}^g_{i 1} \hat{n}^g_{i 2}, \label{eq:H_ef-1d} \\
 \hat{H}^g_0 = & -J_g \sideset{}{_k} \sum [\sigma^x (1 + \cos k) + \sigma^y
 \sin k]_{a b} \, \gd_{k a} \hat{g}_{k b},\nonumber
\end{align}
where $\hat{g}_{k a} = \frac{1}{\sqrt{N_d}} \sum_i \e^{-\ii k x_i} \hat{g}_{i
a}$, ${\bm \sigma}$ are Pauli matrices, momentum $k \in [-\pi, \pi]$ (in units
of inverse lattice spacing $1 / a_0$) is defined in a dimer Brillouin zone (BZ)
with $N_d$ states.
$u_{g g} \! = \! (U^-_{e g})^2 / 4 J_e$ is the strength of intra-dimer $g$-atom
attraction mediated by quantum fluctuations of localized $e$-atoms.
If we associate the site index $a = 1$, $2$ inside a dimer with a
spin-$\frac{1}{2}$ degree of freedom, $\hat{H}^g_0$ contains kinetic energy
with a SOC that arises because any tunneling event ``flips'' pseudospin $a$.

The physical origin of the $p$-wave TSF is especially transparent at weak
coupling $u_{g g} \ll J_g$ and low filling $n^g \ll 1$, when kinetic energy in
\eqref{eq:H_ef-1d} dominates and is diagonalized by states $\hat{f}_{k \tau} =
\frac{1}{\sqrt{2}} \bigl( \hat{g}_{k 1} - \tau \e^{-\ii \frac{k}{2}} \hat{g}_{k
2} \bigr)$ with energies $\epsilon_k = 2 \tau J_g \cos \frac{k}{2}$ ($\tau =
\pm 1$).
Because relevant momenta are small, $\vert k \vert \ll \pi$, it is allowed to
keep only the $\hat{f}_{k, -1} \equiv \hat{f}_k$ mode.
As a result, interactions in \eqref{eq:H_ef-1d} become manifestly $p$-wave:
\begin{equation}
 \hat{H}_{\rm ef} \approx \sideset{}{_k} \sum \epsilon_k \fd_k \hat{f}_k -
 \frac{u_{g g}}{4 N_d} \sideset{}{_{k^\prime k q}} \sum \e^{\ii \frac{q}{2}}
 \fd_{k + q} \fd_{k^\prime - q} \hat{f}_{k^\prime} \hat{f}_k.
 \label{eq:H_ef-1d-f}
\end{equation}
Within the Bogoliubov mean-field theory \cite{suppl} one introduces a
pairing OP $\Delta = -\frac{u_{g g}}{4 N_d} \sum_q \e^{-\ii \frac{q}{2}}
\langle \hat{f}_{-q} \hat{f}_q \rangle$ which parameterizes the single-particle
excitation spectrum $E_k = \sqrt{(\epsilon_k - \mu)^2 + \vert D_k \vert^2}$
with a $p$-wave gap $D_k \approx \ii \, k \Delta$.
$\mu$ is the $g$-atom chemical potential (set by the Fermi energy).
$\Delta$ is a solution of a self-consistency equation $1 = \frac{u_{g g}}{16
N_d} \sum_k \frac{k^2}{E_k}$, $\Delta \approx J_g \e^{-2 \pi J_g / u_{g g}
\sqrt{1 - (\mu / 2 J_g)^2}}$.

\begin{figure}[t]
 \begin{center}
  \includegraphics[width = \columnwidth]{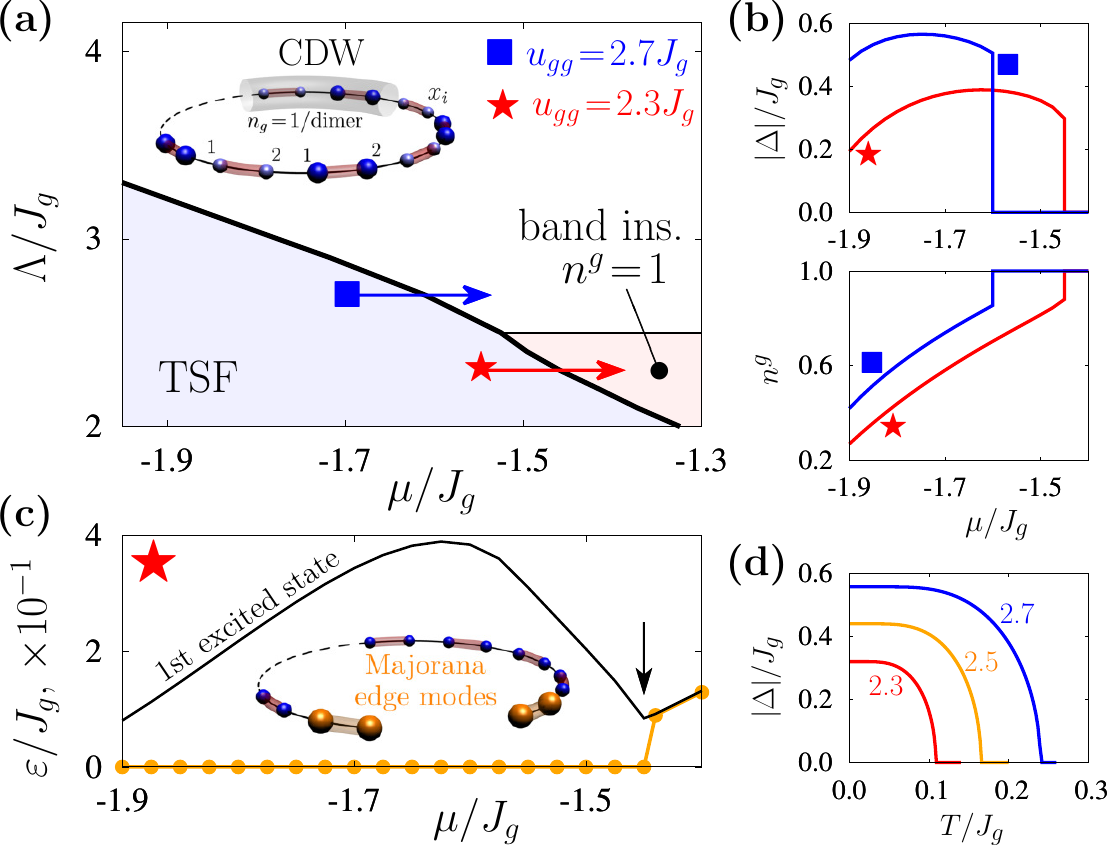}
 \end{center}
 \caption{
  {\bf (a)} Zero-temperature phase diagram of Eq. \eqref{eq:H_ef-1d} computed
  using an unconstrained Hartree-Fock-Bogoliubov mean-field theory in a system
  with $N_d = 100$ dimers and periodic boundary conditions.
  $\mu$ is the $g$-atom chemical potential.
  Thick [thin] lines indicate 1st order transitions between topological
  superfluid (TSF) and insulator states [2nd order transition inside the
  insulating region].
  In the charge-density wave (CDW) state the unit cell has two dimers
  with an average density $n^g = 1$ atom per dimer.
  For small $u_{g g}$ the CDW undergoes a transition to a band insulator with
  a single-dimer unit cell.
  Blue square (red star) corresponds to $u_{g g} / J_g = 2.7$ ($2.3$).
  {\bf (b)} SF gap $\Delta$ and average density $n^g = \frac{1}{N_d} \sum_i
  (n^g_{i 1} + n^g_{i 2})$ plotted along the arrows shown in (a).
  {\bf (c)} Two {\it lowest-magnitude} eigenvalues $\varepsilon$ of the
  BdG Hamiltonian computed in an open chain for $u_{g g} = 2.3 J_g$.
  The order parameters were taken from a converged solution in (a).
  Orange circles indicate Majorana edge modes inside the TSF phase.
  The arrow marks a TSF-band insulator transition.
  {\bf (d)} Gap $\Delta$ as a function of temperature $T$ for $\mu = -1.8
  J_g$, and $u_{g g} / J_g = 2,3$, $2.5$, $2.7$.
  The Boltzmann constant is $k_B = 1$.
 }
 \label{fig:fig2}
\end{figure}

\paragraph*{Stability and topological nature of the SF state.--}

To assess the stability of the $p$-wave SF state beyond the weak coupling
limit, and uncover its topological properties, we compute the phase diagram of
$\hat{H}_{\rm ef}$ within a fully unconstrained Hartree-Fock-Bogoliubov
mean-field approach in real space (explained in Methods).
This variational technique minimizes the grand potential $\bigl \langle
\hat{H}_{\rm ef} - \mu \sum_{i a} \gd_{i a} \hat{g}_{i a} \bigr \rangle$ ($\mu$
is the $g$-atom chemical potential) w.r.t. local OPs $\Delta_i = -u_{g g}
\langle \hat{g}_{i 2} \hat{g}_{i 1} \rangle$, $\xi_i = \langle \gd_{i 2}
\hat{g}_{i 1} \rangle$ and $n^g_{i a} = \langle \gd_{i a} \hat{g}_{i a}
\rangle$, and includes the competition between SF phases with a finite gap
$\Delta_i$ and various inhomogeneous states, e.g. charge-density waves (CDWs),
characterized by a site-dependent $n^g_{ia}$.
The minimization is performed at zero temperature $T = 0$ in a system with
periodic boundary conditions (BCs).
Once the GS is self-consistently determined, we open the chain and diagonalize
the Bogoliubov-de Gennes (BdG) mean-field Hamiltonian {\it with fixed OPs} to
determine edge modes: If a SF phase displays zero-energy (Majorana) modes, we
call it topological \cite{qi-2011-1}.

Fig. \ref{fig:fig2}(a) shows the phase diagram of model \eqref{eq:H_ef-1d} as a
function of chemical potential $\mu$ and interaction $u_{g g}$.
In agreement with the previous section, the SF phase is stable at weak coupling
$u_{g g} < J_g$ and low density, and is characterized by the mixing of singlet
$\langle \hat{g}_{-k, 2} \hat{g}_{k 1} \rangle$ and triplet $\langle
\hat{g}_{-k, a} \hat{g}_{k a} \rangle$ Cooper pair amplitudes.
To better understand this effect, let us consider properties of the model
\eqref{eq:H_ef-1d} under space inversion $I$: $x \to -x$.
In the dimerized lattice, $I = \sigma^x \otimes I_d$, where $I_d$: $k \to -k$
is an inversion acting on the dimer center-of-mass and $\sigma^x$ appears
because $I$ must interchange dimer sites.
The SOC $\hat{H}^g_0$ is invariant under $I$ but {\it manifestly breaks} $I_d$
due to the odd-momentum terms.
In a SF state, Cooper pair wavefunctions inherit this feature and the system
exhibits $p$-wave pairing between same-flavor $g$-atoms $\langle \hat{g}_{-k,
a} \hat{g}_{k a} \rangle \sim k^n$ ($n$ is odd) despite the $s$-wave nature of
interactions in Eq. \eqref{eq:H_ef-1d}.
This situation is similar to singlet-triplet mixing in non-centrosymmetric
superconductors with Rashba SOC \cite{gorkov-2001-1,yip-2002-1,smidman-2017-1}.

When $u_{g g}$ or $\mu$ is increased, the system undergoes a 1st order
transition to a non-SF gapped state with an average density $n^g = 1$ [Fig.
\ref{fig:fig2}(b)].
This phase is a band insulator for small $u_{g g}$, and a CDW with two dimers
per unit cell for strong interactions.
As shown in Fig. \ref{fig:fig2}(c), the SF state is {\it topological} (i.e.
possesses zero-energy edge modes) for all $u_{g g}$ and $\mu$ where it is
stable.
This happens because we used the Hartree-Fock OPs in our variational scheme: if
the minimization were constrained to include only site-independent $\Delta_i$,
one would recover a well-known transition \cite{qi-2011-1} from TSF to a
non-topological SF state.
The latter phase is unstable towards CDW formation and the transition never
happens.

\begin{figure}[t]
 \begin{center}
  \includegraphics[width = \columnwidth]{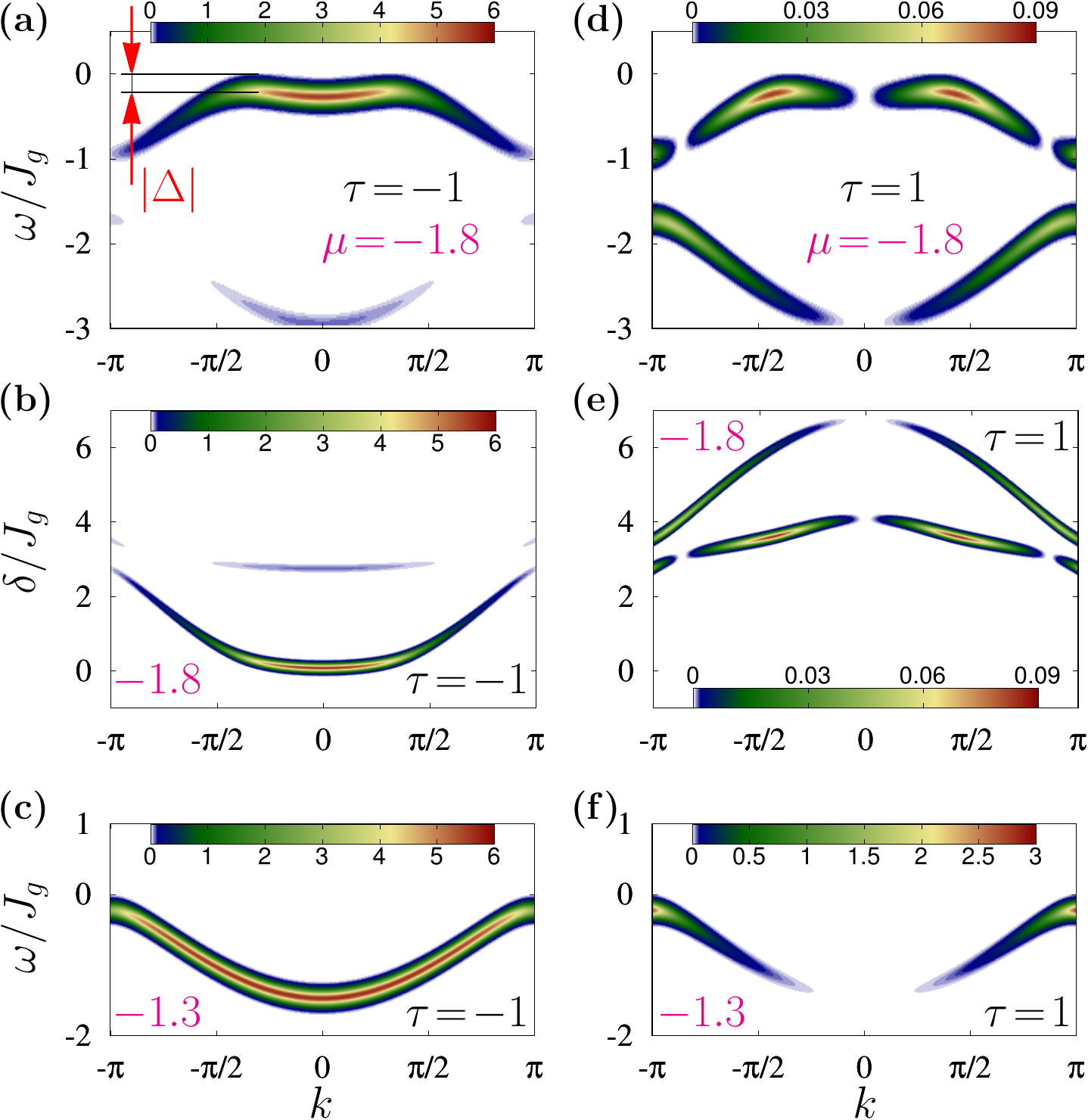}
 \end{center}
 \caption{
  The transfer rate ${\cal R}_\tau (\delta, k)$ (in arbitrary units) of
  $g$-atoms from the TSF with $\mu = -1.8 J_g$, and insulator with $\mu = -1.3
  J_g$ [see Fig. \ref{fig:fig2}(a)].
  These values of $\mu$ are indicated by numbers in magenta.
  The interaction strength is $u_{g g} = 2.3 J_g$.
  For concreteness, in the empty tube we assume $J_e = J_g$ and $\eta = 1$, so
  that $e$-atom band structure is $\epsilon^e_{k \tau} = 2 \tau J_g \cos
  \frac{k}{2}$.
  The left (right) column corresponds to $\tau = \mp 1$.
  {\bf (a)} and {\bf (d)} ${\cal R}_\tau$ as a function of the shifted
  frequency $\omega = \epsilon^e_{k \tau} - \mu - \delta$, which reveals the
  BdG band structure and allows us to extract the SF gap $\Delta$.
  ({\bf b}) and ({\bf e}) The signal in the detuning-momentum plane, as it
  would be measured in a real experiment.
  ({\bf c}) and ({\bf f}) Same as in panels (a) and (d), but inside the band
  insulator with a fully filled {\it single-particle} band $\epsilon_{k, -1}$.
  A weaker signal for $\tau = 1$ is due to Hartree-Fock corrections.
 }
 \label{fig:fig3}
\end{figure}

\begin{figure}[t]
 \begin{center}
  \includegraphics[width = \columnwidth]{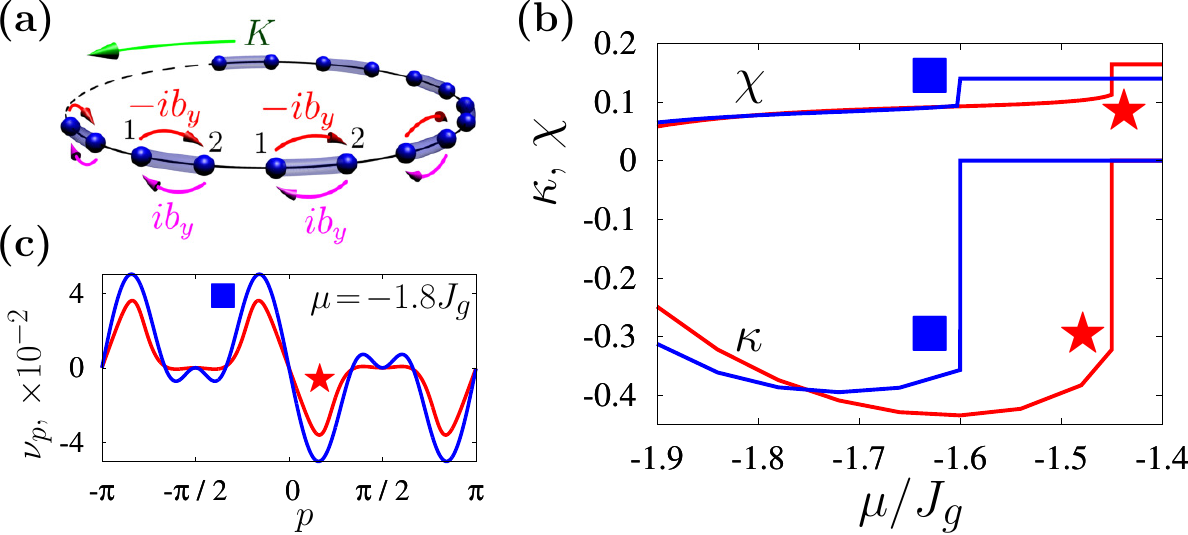}
 \end{center}
 \caption{
  {\bf (a)} Model \eqref{eq:H_ef-1d} with a laser-assisted intra-dimer
  tunneling $\pm \ii \, b_y$ simulating a magnetic field ${\bm B} = b_y {\bm
  e}_y$.
  A supercurrent $\langle \hat{K} \rangle \sim b_y$ is induced in the TSF
  phase.
  {\bf (b)} Magneto-electric response $\kappa = \langle \hat{K} \rangle / b_y$
  and longitudinal susceptibility $\chi = \langle \hat{S}^y \rangle / b_y$ as
  functions of the chemical potential $\mu$.
  {\bf (c)} Momentum distribution asymmetry $\nu_p$ inside the TSF phase for
  $b_y = 0.3 J_g$ and $\mu = -1.8 J_g$.
  In (b) and (c), a red star (blue square) corresponds to $u_{g g} / J_g =
  2.3$ ($2.7$) [cf Fig. \ref{fig:fig2}].
 }
 \label{fig:fig4}
\end{figure}

The phase diagram in Fig. \ref{fig:fig2}(a) remains valid at finite
temperature $T > 0$.
Indeed, as demonstrated in Fig. \ref{fig:fig2}(d) a typical critical
temperature, above which the SF phase disappears, is $T_c \sim 0.1 J_g$ (here
and below we use the units with Boltzmann constant $k_B = 1$).
For $T > T_c$, the system becomes a homogeneous Fermi liquid.

\begin{figure*}[!t]
 \begin{center}
  \includegraphics[width = 1.9 \columnwidth]{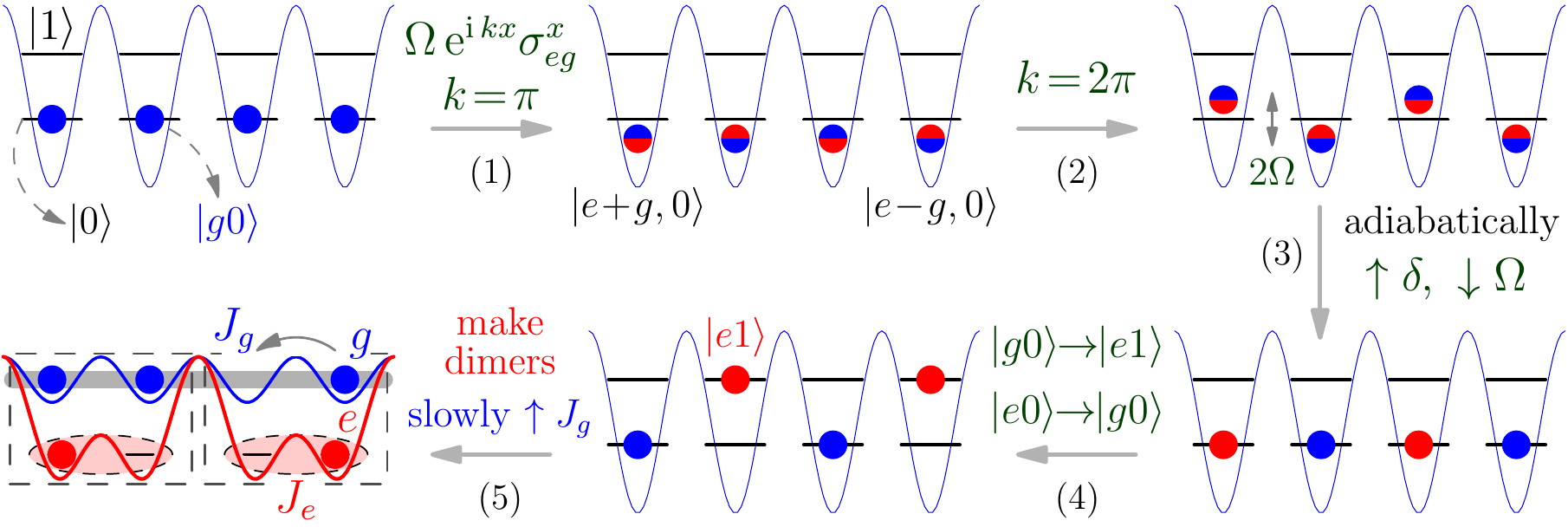}
 \end{center}
 \caption{
  A protocol to prepare the model in Fig. \ref{fig:fig1}(a) and Eq.
  \eqref{eq:H_1d}.
  Blue (red) color indicates $g$ ($e$) atoms.
  Atoms are nuclear-spin polarized and hoppings are quenched until step (5).
  $\vert e (g) \, n \rangle$ means a state with one $e$ ($g$) atom in the
  $n$-th spatial level ($n = 0$ means GS).
  At steps (2) and (3), red-blue (blue-red) circles [from bottom to top]
  correspond to $\vert e \pm g \rangle$ states detuned from the lowest-band
  energy by $\pm \Omega$, respectively.
 }
 \label{fig:fig5}
\end{figure*}

\paragraph*{Detection of the TSF state.--}

The simplest way to validate our theory in cold-atom experiments, is to probe
the $g$-atom attraction via quench dynamics in a {\it normal} state using the
following protocol: {\it (i)} For times $t < 0$, $g$-atoms fill a
non-interacting Fermi sea.
{\it (ii)} At $t = 0$, $J_g$ is switched off (e.g. by increasing the lattice
depth), and $g$-atoms are brought in contact with $e$-atoms, thus allowing them
to experience $e$-$g$ interactions described in Eq. \eqref{eq:H_1d}.
Then, one lets the system evolve for a time $t_0$.
As a result, basis states with doubly occupied dimers accumulate a phase
$-U t_0$, where $U [= -u_{g g}$ in Eq. \eqref{eq:H_ef-1d}] is the induced
$g$-atom interaction.
{\it (iii)} At $t = t_0$, the $e$-atoms are removed, hopping $J_g$ is restored
and the system evolves with a non-interacting Hamiltonian [1st term in
\eqref{eq:H_ef-1d}].
The sign of $U$ can be determined by measuring an average number of doubly
occupied dimers: $n_2 (t) = \langle \psi (t) \vert \sum_i \hat{n}^g_{i 1}
\hat{n}^g_{i 2} \vert \psi (t) \rangle$ in the state $\vert \psi (t) \rangle$
of the system at time $t$.
For short evolution times when $\vert U \vert t_0 \ll 1$ and $t - t_0 \ll 1 /
J_g$, $n_2 (t) = n_2 (0) - \zeta \, U \cdot (t - t_0)$ with $\zeta > 0$.
Hence, the number of double occupancies decreases (increases) for the repulsive
(attractive) interaction $U$ (see Methods for a complete derivation of this
result).

The spectrum of Bogoliubov excitations can be probed by momentum-resolved
spectroscopy \cite{stewart-2008-1,dao-2009-1}.
Let us assume that one tube in Fig. \ref{fig:fig1}(a) contains no atoms and is
detuned relative to its neighboring tubes.
A laser with a wavevector along the $x$-axis, Rabi frequency $\Omega$, detuned
by $\delta$ from the atomic $e$-$g$ transition, transfers $g$-atoms from the SF
phase to $e$-states in the empty tube (the transfer happens along the $y$-axis
and does not change an atom's $x$-position along the tube).
The $e$-lattice depth in that tube has been reduced to make flat bands at $\pm
J_e$ [see Fig. \ref{fig:fig1}(b)] dispersive with a band structure
$\epsilon^e_{k \tau} = \tau J_e \sqrt{1 + \eta^2 + 2 \eta \cos k}$ where $\tau
= \pm 1$ and $\eta J_e$ is the inter-dimer hopping.
The $g$-atom transfer rate to an $e$-band $\epsilon^e_{k \tau}$, ${\cal R}_\tau
(\delta, k)$, can be written in terms of the spectral density ${\cal A}_{a b}
(\omega, k)$ of the single-particle normal Green function \cite{dao-2009-1}:
${\cal R}_\tau (\delta, k) = \frac{\Omega^2}{2} \bigl[ {\cal A}_{1 1} (\omega,
k) + {\cal A}_{2 2} (\omega, k) - 2 \tau \frac{{\rm Re} \, [(1 + \eta \e^{\ii
k}) {\cal A}_{1 2} (\omega, k)]}{\sqrt{1 + \eta^2 + 2 \eta \cos k}} \bigr]$
with $\omega = \epsilon^e_{k \tau} - \mu - \delta$, ${\cal A}_{a b} (\omega, k)
= \ii \, f (\omega) [\langle \hat{g}_{k a} \gd_{k b} \rangle_{\omega + \ii 0} -
\langle \hat{g}_{k a} \gd_{k b} \rangle_{\omega - \ii 0}]$.
$f (\omega) = (\e^{\omega / T} + 1)^{-1}$ is the Fermi function at temperature
$T$, and $\ii 0$ is an infinitesimal imaginary number (see Methods).

The representative signal ${\cal R}$ is shown in Fig. \ref{fig:fig3}.
Its maximum (for a fixed $k$) occurs when $\omega$ coincides with the highest
occupied BdG energy state.
Therefore, one can map out the BdG band structure and extract the excitation
gap $\Delta$ [see panels (a) and (d)].
In Fig. \ref{fig:fig3}(b) and (e) we plot the same signal as a function of the
bare laser detuning $\delta$, as it would be observed in an experiment.
This spectroscopy technique can also be used to probe the insulating phase in
Fig. \ref{fig:fig1}(a).
The transfer rate inside the band insulator regime is shown in Fig.
\ref{fig:fig3}(c) and (f).
In this case, the excitation spectrum is again gapped, but as opposed to the SF
state, this gap exists because all {\it single-particle} states below the Fermi
level are filled, and not due to fermion pairing.
Finally, we note that the signal with $\tau = -1$ [panels (a) -- (c)] is
significantly stronger than the one with $\tau = 1$ [(d) -- (f)], which
highlights the validity of the effective model \eqref{eq:H_ef-1d-f}.

Due to the SOC inherent in Eq. \eqref{eq:H_ef-1d}, the TSF state has remarkable
features that set it apart from a usual $s$-wave SF and can be used as its
``fingerprint''.
Perhaps its most revealing property is an analog of the spin-galvanic effect,
when an applied Zeeman magnetic field induces a bulk supercurrent
\cite{yip-2002-1,ojanen-2012-1}.
Since the pseudospin degrees of freedom in $\hat{H}_{\rm ef}$ correspond to a
site index inside the unit cell, this ``field'' must couple to the motion of
$g$-atoms and can be implemented as a laser-assisted tunneling within dimers
\cite{goldman-2014-1,galitski-2013-1}.
This hopping can be set to have an arbitrary phase $\theta$, but we focus on
the case $\theta = \frac{\pi}{2}$, i.e. consider a perturbation: $\delta
\hat{H}_{\rm ef} = -b_y \sum_{i, \, a b} \sigma^y_{a b} \, \gd_{i a}
\hat{g}_{i b}$ and compute supercurrent response $\langle \hat{K} \rangle =
\kappa \, b_y$ [Fig. \ref{fig:fig4}(a)].
We can anticipate this response based on pure symmetry arguments: as
explained previously in this section, the SOC, being odd in momentum, breaks
the space inversion symmetry in the dimer lattice.
On the other hand, the synthetic Zeeman term $\delta H_{\rm ef}$ with $b_y \neq
0$ violates time-reversal symmetry (see discussion in Methods).
Breaking of these two symmetries is a necessary condition to stabilize a state
with a non-zero current in the system.

The operator $\hat{K}$ is a single-particle mass current, obtained by varying
the Hamiltonian $\hat{H}_b = \hat{H}^g_0 + \delta \hat{H}_{\rm ef}$ [cf Eq.
\eqref{eq:H_ef-1d}] w. r. t. the flux $\varphi$ piercing the ring: $\hat{K} =
\delta \hat{H}_b / \delta \varphi \bigl \vert_{\varphi = 0}$.
This flux enters via a phase factor $\e^{\ii \varphi}$ on each physical link in
the lattice with one-site unit cell, but in the dimerized lattice one must
replace $\hat{g}_{i 1} \to \e^{2 \ii \varphi x_i} \hat{g}_{i 1}$ and
$\hat{g}_{i 2} \to \e^{\ii (2 x_i + 1) \varphi} \hat{g}_{i 2}$, because the
position $x$ in a non-dimerized lattice is $x = 2 x_i + a - 1 = 0, \ldots, 2
N_d - 1$ and $\hat{g}_x = \hat{g}_{i a}$.
We obtain $\hat{K} = \frac{1}{N_d} \sum_k \bigl[ \sigma^y (1 - \cos k) +
\sigma^x \bigl( \sin k - b_y / J_g \bigr) \bigr]_{a b} \, \gd_{k a} g_{k b}$.
In Fig. \ref{fig:fig4}(b) we show the magneto-electric coefficient $\kappa$ as
a function of the chemical potential for several interaction strengths.
Remarkably, $\kappa$ exists {\it only inside} the SF phase and vanishes across
the TSF-insulator transition.

Due to the $p$-wave nature of the SF phase, the synthetic field $b_y$ induces a
pseudospin polarization $\langle \hat{S}_y \rangle = \chi b_y$, where
$\hat{\bm S} = \frac{1}{2 N_d}\sum_i {\bm \sigma}_{a b} \gd_{i a} \hat{g}_{i
b}$.
The susceptibility $\chi$ and magneto-electric coefficient $\kappa$ can be
related in the weak-coupling dilute limit $u_{g g} \ll J_g$, $n^g \ll 1$ at $T
= 0$ \cite{gorkov-2001-1}.
Indeed, as explained in Methods, similar calculations that led to Eq.
\eqref{eq:H_ef-1d-f} yield $\kappa = -4 \chi = -\frac{1}{2 J_g N_d} \sum_k
\bigl[ 1 - \frac{\epsilon_k - \mu}{E_k} \bigr] \cos \frac{k}{2}$ with
$\epsilon_k$ and $E_k$ defined after Eq. \eqref{eq:H_ef-1d-f}.
Fig. \ref{fig:fig4}(b) shows $\chi$ and $\kappa$ as functions of the chemical
potential $\mu$ across the TSF-insulator transition.
In cold-atom experiments it is possible to measure $\chi$, for example by a
Ramsey-type protocol \cite{bromley-2017-1}: Assuming that the system is in its
GS $\vert \psi_0 \rangle$ (with $b_y \neq 0$), at $t = 0$ we quench the
Hamiltonian from \eqref{eq:H_ef-1d} to $\hat{H}_x = J_g \hat{S}_x$ e.g. by
making the intra-dimer $g$-atom tunneling dominant and let the system evolve
for a time $t_0 = \frac{\pi}{2 J_g}$.
As a result, the state becomes $\vert \psi \rangle = \e^{-\ii \frac{\pi}{2}
\hat{S}_x} \vert \psi_0 \rangle$.
Now we measure the difference in populations on two sites of a dimer, i.e.
$\langle \psi \vert \hat{S}_z \vert \psi \rangle$.
Because $\e^{\ii \frac{\pi}{2} \hat{S}_x} \hat{S}_z \e^{-\ii \frac{\pi}{2}
\hat{S}_x} = \hat{S}_y$, the above protocol yields $\langle \psi_0 \vert
\hat{S}_y \vert \psi_0 \rangle$ and can be used to obtain $\chi$.

Another physical effect induced by $\delta \hat{H}_{\rm ef}$ is an asymmetry of
the $g$-atom momentum distribution which can be detected in time-of-flight
experiments \cite{mancini-2015-1}.
Because these measurements involve crystal momentum $p$ in the BZ of a
{\it single-site} unit cell, we need to compute $\nu_p = \frac{J_g}{b_y}
\langle \gd_p \hat{g}_p - \gd_{-p} \hat{g}_{-p} \rangle$ with $\hat{g}_p \equiv
\frac{1}{\sqrt{2 N_d}} \sum_x \e^{-\ii p x} \hat{g}_x = \frac{1}{\sqrt{2 N_d}}
\sum_{i a} \e^{-2 \ii p x_i - \ii p (a - 1)} \hat{g}_{i a} = \frac{1}{\sqrt{2}}
\bigl( g_{2 p, 1} + \e^{-\ii p} g_{2 p, 2} \bigr)$.
Fig. \ref{fig:fig4}(c) shows $\nu_p$ computed in the TSF phase of Fig.
\ref{fig:fig2}(a).
For comparison, in a non-SF system, $\nu_p \sim \delta_{k, k_F} -
\delta_{k, -k_F}$ ($k_F$ is the Fermi momentum).
SF correlations destroy Fermi points and lead to a finite asymmetry even away
from $k = \pm k_F$.

\paragraph*{Preparation of the ${}^{87} {\rm Sr}$ lattice clock.--}

The system in Fig. \ref{fig:fig1}(a) and Eq. \eqref{eq:H_1d} can be realized
with AEAs, such as fermionic ${}^{87} {\rm Sr}$, using steps shown in Fig.
\ref{fig:fig5}:
{\bf Step (0)} We start with a nuclear-spin polarized $g$-atom band insulator
in a deep magic-wave lattice (in which $e$ and $g$ atoms experience equal light
shifts and therefore same trapping potential \cite{ye-2008-1}) with suppressed
tunneling.
{\bf Step (1)} The system is irradiated by a laser that adiabatically applies a
staggered synthetic gauge field \cite{cooper-2015-1,wall-2016-1} with
wavevector $k = \pi$, Rabi frequency $\Omega$ and a detuning from the $e$-$g$
transition $\delta$.
In the $e$-$g$ basis, the single-atom Hamiltonian at the lattice site $j$ is
$\hat{H}_j = \frac{1}{2} \bigl[ (-1)^j \Omega \sigma^x - \delta \sigma^z
\bigr]$.
Here the Pauli matrices ${\bm \sigma}$ act on the local $g$-$e$ basis.
As the detuning is decreased to zero, the Rabi frequency is simultaneously
ramped up, thus adiabatically preparing atoms in their local GS
$\frac{1}{\sqrt{2}} \vert e \pm g \rangle_j$.
{\bf Step (2)} The laser wavevector is quenched from $k = \pi$ to $2 \pi$,
making half of the local $e$-$g$ mixtures excited states.
{\bf Step (3)} $\delta$ is adiabatically increased, while $\Omega$ is
decreased.
As a result, excited (GS) coherent $e$-$g$ superpositions are transferred to
$e$ ($g$) states.
{\bf Step (4)} A laser is used to directly transfer GS $e$-atoms [$\vert e \, 0
\rangle$] to $g$-atoms [$\vert g \, 0 \rangle$] and $\vert g \, 0 \rangle$ to
excited $e$-atoms [$\vert e \, 1 \rangle$], where $\vert e (g) \, n \rangle$
indicates an $e$ ($g$) atom in $n$-th lattice band.
{\bf Step (5)} We adiabatically enable hoppings by decreasing the
magic-wavelength lattice depth.
Simultaneously, we create a dimerized $e$-superlattice by ramping up a
potential experienced only by $e$-atoms \cite{safronova-2015-1} at twice the
periodicity of the magic lattice, and transfer the states $\vert e \, 1
\rangle$ to excited antisymmetric motional states in each double-well, in order
to satisfy the requirement $\vert J_e \vert > \vert J_g \vert$.

After the last step, $g$-atoms form a band-insulator with $n^g = 1$ (two atoms
per dimer).
Their filling can be controlled spectroscopically by removing atoms from
$k$-states near band edges with a laser which drives a narrow transition whose
detuning is adiabatically changed to scan the conduction band and access atoms
deeper in the Fermi sea.
This can be visualized as an adiabatic injection of holes in the presence of
$e$-atom background.
The newly added holes form Cooper pairs, thus building up a SF state.

\paragraph*{Discussion.--}

Topological superfluidity in Fermi liquids is intimately related to the
coupling between particles' spin and orbital motion.
Unfortunately, in most systems this crucial ingredient is absent or too weak to
yield a measurable topological structure of SF phases.
In the present work we discussed a mechanism that bypasses this ``rule'' and
allows a {\it coexistence of a strong spin-orbit coupling and pairing
correlations}.
The key ingredient in our theory is the lattice modulations that host localized
degrees of freedom and play a dual role.
On the one hand, quantum fluctuations of localized fermions stabilize SF states
in the itinerant channel, {\it even when the bare interactions are repulsive}.
On the other hand, the modulations enlarge the lattice unit cell and lead to an
emergent {\it odd in momentum} SOC in the conduction band.
A combination of these effects always results in a topologically non-trivial SF
state in a number-conserving system with potential emergence of Majorana modes.
We illustrated the above mechanism by studying a system of spinless fermions in
a quasi-1D lattice with a dimerized structure and showed how one can observe
this physics in a quantum simulator with AEAs with a variety of probes,
including momentum-resolved spectroscopy and an analog of the spin-galvanic
effect, i.e. a magneto-electric phenomenon that can be used to detect a SF
phase with broken inversion symmetry.

The analysis presented above can be easily extended beyond 1D.
In particular, in a 2D system where $g$-atoms propagate in a square lattice,
and $e$-atoms are localized inside square plaquettes, similar arguments show
that quantum fluctuations of the $e$-atoms stabilize a $p_x + \ii \, p_y$ SF
state of the $g$-species \cite{suppl}.

Although our presentation illustrates main ideas behind this emergent
phenomenon using mean-field approximations, the topological nature of the SF
state remains intact beyond mean-field.
We confirmed this by performing exact diagonalization in a single tube and
verifying that the GS realizes a fermionic parity switch \cite{ortiz-2014-1}
for all fermion fillings \cite{suppl}.

\paragraph*{Acknowledgments.--}

We thank Daniel Agterberg, Victor Gurarie, Colin Kennedy, Johannes
Schachenmayer, Dmitry Solenov, and Ilya Vekhter for illuminating discussions.
This work was supported by NSF (PHY-1211914, PHY-1521080 and
JILA-PFC-PHY-1734006), AFOSR FA9550-13-1-0086, AFOSR-MURI Advanced Quantum
Materials, NIST and DARPA W911NF-16-1-0576 through ARO.

\paragraph*{Competing financial interests.--}

The authors declare no competing financial interests.

\end{document}


\title{Methods and Supplementary material}
\maketitle

\section{Methods}
\renewcommand{\theequation}{M\arabic{equation}}

\subsection*{Derivation of the $g$-atom effective Hamiltonian (2)}

We focus on the regime $J_e \gg J_g, \, U^-_{e g}$ when the interaction term
$\hat{H}_{\rm int} = U^-_{e g} \sum_{i a} \hat{n}^g_{i a} \hat{n}^e_{i a}$ and
$g$-atom kinetic energy $\hat{H}^g_0$ in Eq. (1) can be treated as a small
correction to the kinetic energy of $e$-atoms $\hat{H}^e_0 = -J_e
\sideset{}{_i} \sum (\ed_{i 1} \hat{e}_{i 2} + {\rm h.c.})$, and use
perturbation theory to compute the effective low-energy model (2).
The zero-order subspace is spanned by the states $\vert \Psi_{e g} \rangle =
\prod_i \vert \lambda \rangle_i \otimes \vert \Psi_g \rangle$.
Because $J_g \ll J_e$, we consider this subspace degenerate for all $\vert
\Psi_g \rangle$.
To determine 1st (2nd) order corrections to the wavefunctions (energy
eigenvalues), we employ the Schrieffer-Wolff transformation (SWT) method
which systematically removes {\it off-diagonal} matrix elements of
$\hat{H}_{\rm int}$ between states $\vert \Psi_{e g} \rangle$ to each order in
$U^-_{e g}$ \cite{bir-1974-1}: $\hat{H} = \hat{H}^e_0 + \hat{H}_{\rm int} \to
\e^{\cal \hat{S}} \hat{H} \e^{-{\cal \hat{S}}}$ with ${\cal \hat{S}} = {\cal
\hat{S}}_1 + {\cal \hat{S}}_2 + \ldots$ being an anti-hermitian generator of
the SWT and ${\cal \hat{S}}_n \sim (U^-_{e g})^n$.
Notice that $\hat{H}^g_0$ acts only on the $g$-atom states $\vert \Psi_g
\rangle$ and contains only {\it diagonal} matrix elements, thus representing a
1st order correction.
On the other hand, $\hat{H}_{\rm int}$ has both off-diagonal and diagonal
elements.
The latter are proportional to $\sum_{i a} \hat{n}^g_{i a}$, i.e. they only
correct the $g$-atom chemical potential and can be omitted.

To the first order in $U^-_{e g}$, ${\cal \hat{S}}$ is given by:
\begin{displaymath}
 {\cal \hat{S}}_1 = U^-_{e g} \sum_{i a, \, E^\prime_e E_e}
 \frac{\hat{P}_{E^\prime_e} \, \widetilde{\hat{n}^e}_{i a} \, \hat{P}_{E_e}}
 {E^\prime_e - E_e} \, \hat{n}^g_{i a}.
\end{displaymath}
where $\hat{P}_{E_e}$ is the projector onto the state $\prod_i \vert \lambda
\rangle_i$ of the $e$-subsystem with an energy $E_e$, and
$\widetilde{\hat{n}^e}_{i a}$ is an off-diagonal part of the matrix ${}_i
\langle \lambda^\prime \vert \hat{n}^e_{i a} \vert \lambda \rangle_i =
\frac{1}{2} \delta_{\lambda^\prime \lambda} + \frac{1}{2}
\sigma^x_{\lambda^\prime \lambda} (\delta_{a 1} - \delta_{a 2})$ ($\delta_{a
b}$ is a Kronecker delta-symbol).

The generator ${\cal \hat{S}}_1$ yields a 2nd order in $U^-_{e g}$ correction
$\hat{H}_2 = \frac{1}{2} \bigl[ {\cal \hat{S}}_1, \hat{H}_{\rm int} \bigr]$
which operates in the degenerate manifold $\lbrace \vert \Psi_{e g} \rangle
\rbrace$:
\begin{align}
 \hat{H}_2 = & \frac{(U^-_{e g})^2}{2} \sum_{\substack{i a b\\ E^\prime_e E_e}}
 {}_i \bigl \langle \lambda \bigl \vert \frac{\hat{P}_{E^\prime_e} \,
 \widetilde{\hat{n}^e}_{i a} \, \hat{P}_{E_e}}{E^\prime_e - E_e}
 \widetilde{\hat{n}^e}_{i b} \bigr \vert \lambda \bigr \rangle_i
 \hat{n}^g_{i a} \hat{n}^g_{i b} + {\rm h.c.} \nonumber \\
 & = \lambda u_{g g} \sideset{}{_i} \sum \hat{n}^g_{i 1} \hat{n}^g_{i 2}
 \nonumber
\end{align}
with $u_{g g} = (U^-_{e g})^2 / 4 J_e$ and $\lambda = \pm 1$.
The $g$-atom interaction can be controlled by preparing the $e$-subsystem in a
particular local state: it is repulsive (attractive) for a (anti-) symmetric
$e$-state $\lambda = +1$ ($\lambda = -1$).

\subsection*{Unconstrained Hartree-Fock-Bogoliubov theory in position space}

In the present section, we summarize the fully unconstrained
Hartree-Fock-Bogoliubov (HFB) mean-field approach \cite{blaizot-1986-1} which
we used to compute the phase diagram in Fig. 2.
This variational technique treats on an equal footing SF and normal phases,
including various insulating states that break translational symmetry, and
hence provides an additional check of robustness of the TSF phase.
Our method amounts to the following linearization of the four-fermion
interaction:
\begin{align}
 \hat{H}_{\rm int} = & - \! u_{g g} \! \sideset{}{_i} \sum \! \gd_{i 1}
 \gd_{i 2} \hat{g}_{i 2} \hat{g}_{i 1} \! \approx - \! u_{g g} \!
 \sideset{}{_i} \sum \! \bigl[ \hat{h}_H \! + \! \hat{h}_F \! + \! \hat{h}_B
 \bigr]; \nonumber \\
 \hat{h}_{H} = & \langle \hat{n}^g_{i 1} \rangle \gd_{i 2} \hat{g}_{i 2} +
 \langle \hat{n}^g_{i 2} \rangle \gd_{i 1} \hat{g}_{i 1}; \nonumber \\
 \hat{h}_F = & -\langle \gd_{i 1} \hat{g}_{i 2} \rangle \gd_{i 2} \hat{g}_{i 1}
 - {\rm h.c.} = \xi_i \gd_{i 1} \hat{g}_{i 2} + {\rm h.c.}; \nonumber \\
 \hat{h}_B = & \langle \hat{g}_{i 2} \hat{g}_{i 1} \rangle \gd_{i 1} \gd_{i 2}
 + {\rm h.c.} = -\frac{\Delta_i}{u_{g g}} \gd_{i 1} \gd_{i 2} - {\rm h.c.},
 \label{eq:H_int_HFB}
\end{align}
where $\hat{h}_{H, F, B}$ refer to Hartree, Fock and Bogoliubov contributions
with (in general complex) OPs $n^g_{i a} = \langle \gd_{i a} \hat{g}_{i a}
\rangle$, $\xi_i = -\langle \gd_{i 2} \hat{g}_{i 1} \rangle$, and $\Delta_i =
-u_{g g} \langle \hat{g}_{i 2}\hat{g}_{i 1} \rangle$.
The full HFB mean-field Hamiltonian includes $\hat{H}_{\rm int}$ and the
$g$-atom kinetic energy $\hat{H}^g_0$ [see Eq. (2)], and can be written as:
\begin{displaymath}
 \hat{H}_{\rm HFB} = \frac{1}{2} \sum_{\alpha \beta}
 \begin{pmatrix}
  \gd_\alpha & \hat{g}_\alpha
 \end{pmatrix}
 \underbrace{
  \begin{pmatrix}
   X_{\alpha \beta} & D_{\alpha \beta} \\
   -D^*_{\alpha \beta} & -X^*_{\alpha \beta}
  \end{pmatrix}
 }_{{\cal H}_{\alpha \beta}}
 \begin{pmatrix}
  \hat{g}_\beta \\
  \gd_\beta
 \end{pmatrix}
\end{displaymath}
with composite indices $\alpha = \lbrace i a \rbrace$ and $\beta = \lbrace j b
\rbrace$, and $2 N_d \times 2 N_d$ dimensional matrices $D_{\alpha \beta} = \ii
\sigma^y_{\alpha \beta} \Delta_i \delta_{i j}$ and
\begin{align}
 X_{\alpha \beta} \! = \! X_{i a, \, j b} \! = & -J_g \bigl[ \delta_{i j}
 \sigma^x_{a b} + \bigl( \delta_{a 2}^{b 1} \delta_{j, i + 1} +
 \delta_{a 1}^{b 2} \delta_{j, i - 1} \bigr) \bigr] \nonumber \\
 & -\mu \delta_{i j} \delta_{a b} - u_{g g} \delta_{i j}
 \begin{pmatrix}
  n^g_{i 2} & \xi_i \\
  \xi^*_i & n^g_{i 1}
 \end{pmatrix}
 _{a b}. \nonumber
\end{align}
The kernel ${\cal H}_{\alpha \beta}$ has the property $\tau^x {\cal H}^* \tau^x
= -{\cal H}$ with $\tau^x = (\sigma^x \otimes \delta_{\alpha \beta})$, which
guarantees that its spectrum is even w.r.t. zero energy. We shall enumerate its
positive eigenvalues by $\nu = 1, \ldots, n_\nu$ with $n_\nu = 2 N_d$.
$\hat{H}_{\rm HFB}$ can be diagonalized by a Bogoliubov transformation
\begin{displaymath}
 \begin{pmatrix}
  \hat{g}_\alpha \\
  \gd_\alpha
 \end{pmatrix}
 =
 \sum_\nu^{n_\nu} \bigl[ \psi^\nu \hat{\gamma}_\nu + (\tau^x \psi^\nu)^*
 \hat{\gamma}^\dag_\nu \bigr],
\end{displaymath}
where $\hat{\gamma}_\nu$ are new fermionic modes and $\psi^\nu$ is an
eigenstate of ${\cal H}$ corresponding to an energy $E_\nu \geqslant 0$.
These states obey a completeness relation $\sum_\nu^{n_\nu} \bigl[ \psi^{*
\nu}_p \psi^\nu_q + (\tau^x \psi^\nu)_p (\tau^x \psi^\nu)^*_q \bigr] =
\delta_{p q}$ with $p$ and $q = 1, \ldots, 4 N_d$.

Using the above quasiparticle modes, we can write down the self-consistency
relations:
\begin{align}
 \Delta_i = & -\frac{u_{g g}}{2} \sideset{}{^{n_\nu}_\nu} \sum \bigl[
 \psi^{* \nu}_{(i 1) + 2 N_d} \psi^\nu_{(i 2)} - \psi^{* \nu}_{(i 2) + 2 N_d}
 \psi^\nu_{(i 1)} \bigr] \phi_\nu, \nonumber \\
 n^g_{i a} = & \frac{1}{2} - \frac{1}{2} \sideset{}{^{n_\nu}_\nu} \sum \bigl[
 \vert \psi^\nu_{(i a)} \vert^2 - \vert \psi^\nu_{(i a) + 2 N_d} \vert^2 \bigr]
 \phi_\nu, \nonumber \\
 \xi_i = & \frac{1}{2} \sideset{}{^{n_\nu}_\nu} \sum \bigl[
 \psi^{* \nu}_{(i 2)} \psi^\nu_{(i 1)} - \psi^{* \nu}_{(i 1) + 2 N_d}
 \psi^\nu_{(i 2) + 2 N_d} \bigr] \phi_\nu, \nonumber
\end{align}
where $\phi_\nu = 1 - 2 f_\nu$ and $f_\nu = \langle \hat{\gamma}^\dag_\nu
\hat{\gamma}_\nu \rangle = 1 / (\e^{E_\nu / T} + 1)$.

The quantum phase diagram (Fig. 2) was computed by minimizing the grand
potential ${\cal G} = \langle \hat H_{\rm HFB} \rangle$ (notice, $\hat{H}_{\rm
HFB}$ already includes the term $-\mu \sum_{i a} \hat{n}^g_{i a}$):
\begin{align}
 \frac{\cal G}{N_d} = -\mu & - \frac{1}{N_d} \sum_i \biggl[ \frac{\vert
 \Delta_i \vert^2}{u_{g g}} + u_{g g} \bigl( n^g_{i 1} n^g_{i 2} - \vert \xi_i
 \vert^2 \bigr) \biggr] \nonumber \\
 - \frac{1}{2 N_d} & \sum_{\alpha \beta} X_{\alpha \beta}(u_{g g} \! = \! 0) \!
 \sum_\nu^{n_\nu} \! \bigl[ \psi^{* \nu}_\alpha \psi^\nu_\beta -
 \psi^{* \nu}_{\beta + 2 N_d} \psi^\nu_{\alpha + 2 N_d}\bigr]. \nonumber
\end{align}

\subsection*{Probing $g$-atom attraction}

Here we present a simple derivation of the time dependence of the number of
doubly occupied dimers after the quench protocol described in the main text.
Assume that the $g$-$g$ interactions are Hubbard-like: $\hat{H}_{\rm int} = U
\hat{V} = U \sum_i \hat{n}^g_{i 1} \hat{n}^g_{i 2}$, where $U$ denotes the
interaction strength whose sign is unknown and needs to be determined (in the
main text, $U = -u_{g g} < 0$).
The operator $\hat{V}$ is a projector onto states with only doubly-occupied
dimers.
Our task is to compute $n_2 (t) = \langle \psi (t) \vert \hat{V} \vert \psi (t)
\rangle$, where $\vert \psi (t) \rangle$ is the time-dependent state of the
system.

At time $t = 0$ $g$-atoms fill the Fermi sea $\vert \psi_0 \rangle = \vert \psi
(t = 0) \rangle = \prod _{k < k_F} \fd_{k, -1} \vac$, where $\vac$ is the
vacuum state without particles and $k_F$ is the Fermi momentum.
The wavefunction at $t = t_0$ is $\vert \psi (t_0) \rangle = \e^{-\ii \varphi
\hat{V}} \vert \psi_0 \rangle$ with $\varphi = U t_0$, while for $t > t_0$ it
is given by $\vert \psi (t > t_0) \rangle = \e^{-\ii \hat{H}_0^g (t - t_0)}
\vert \psi (t_0) \rangle$.
Therefore,
\begin{displaymath}
 n_2 (t > t_0) = \langle \psi_0 \vert \e^{\ii \varphi \hat{V}}
 \e^{\ii \hat{H}_0^g (t - t_0)} \hat{V} \e^{-\ii \hat{H}_0^g (t - t_0)}
 \e^{-\ii \varphi \hat{V}} \vert \psi_0 \rangle.
\end{displaymath}
Consider a short time $t_0 \ll 1 / U$ when $\varphi \ll 1$ and $n_2 (t) -
\langle \psi_0 \vert \hat{V} \vert \psi_0 \rangle \approx \ii \varphi \langle
\psi_0 \vert \bigl[ \hat{V}, \e^{\ii \hat{H}_0^g \tau} \hat{V} \e^{-\ii
\hat{H}_0^g \tau} \bigr] \vert \psi_0 \rangle$.
If in addition we limit ourselves to short-time dynamics, $t - t_0 \ll 1 /
J_g$, then $\e^{\ii \hat{H}_0^g (t - t_0)} \hat{V} \e^{-\ii \hat{H}_0^g (t -
t_0)} \approx \hat V + \ii (t - t_0) [\hat{H}_0^g, \hat{V}]$, and the final
expression for $n_2 (t)$ is:
\begin{displaymath}
 n_2 (t) = \langle \hat{V} \rangle + \varphi (t - t_0) \langle [\hat{V},
 [\hat{V}, \hat{H}_0^g]] \rangle = \langle \hat{V} \rangle - \kappa \varphi
 (t - t_0).
\end{displaymath}
Here $\langle \ldots \rangle = \langle \psi_0 \vert \ldots \vert \psi_0
\rangle$ and $\kappa = 2 \bigl[ \langle \hat{V} \hat{H}_0^g \hat{V} \rangle -
E_0 \langle \hat{V}^2 \rangle \bigl]$ ($E_0$ is the GS energy).
It is easy to see that $\kappa > 0$.
Indeed, introducing a normalized wavefunction $\vert \psi^V_0 \rangle = \hat{V}
\vert \psi_0 \rangle / \langle \hat{V}^2 \rangle$, we have $\kappa = 2 \langle
\hat{V}^2 \rangle \bigl[ \langle \psi^V_0 \vert \hat{H}_0^g \vert \psi^V_0
\rangle - E_0 \bigr]$, which is clearly positive.

\subsection*{Momentum-resolved spectroscopy signal}

Here we derive the expression for ${\cal R}_\tau (\delta, k)$ used in the main
text.
Our approach almost exactly follows Ref. \onlinecite{dao-2009-1}.
We use units such that $\hbar = k_B = 1$.

The transfer of $g$-atoms from the SF phase to $e$-states in an empty tube is
governed by the operator:
\begin{displaymath}
 \hat{V}_L = \Omega \sideset{}{_{i a}} \sum \hat{q}^\dag_{i a} \hat{g}_{i a},
\end{displaymath}
where $\Omega$ is the Rabi frequency and $\hat{q}^\dag_{i a}$ creates an
$e$-atom in well $a$ of the $i$-th dimer in the auxiliary (empty) tube.
The initial state of the system contains $N_g$ $g$-atoms, $\vert \psi_{\rm in}
\rangle = \vert \psi^{i, g}_{N_g} \rangle \otimes \vac$ ($\vac$ is an empty
state of the auxiliary tube).
Its energy is ${\cal E}_{\rm in} = E^g_i + \mu N_g + \omega_0$, where
$\omega_0$ is the laser frequency (note that we work in the grand-canonical
ensemble, so $E^g_i$ already includes the chemical potential offset).
In the final state, there is a single $e$-atom with momentum $k$ and band index
$\tau$ in the auxiliary tube: $\vert \psi_{\rm fi} \rangle = \vert \psi^{f,
g}_{N_g - 1} \rangle \otimes \hat{f}^\dag_{k \tau} \vac$ ($\vert \psi^{f,
g}_{N_g - 1} \rangle$ is an intermediate state of $g$-atoms with $N_g - 1$
particles).
This state has an energy ${\cal E}_{\rm fi} = E^g_f + \mu (N_g - 1) + \nu_{e g}
+ \epsilon^e_{k \tau}$, where $\nu_{e g}$ is the $e$-$g$ transition frequency
($\sim 10^{14}$ Hz for ${}^{87} {\rm Sr}$).
Because $\nu_{e g}$ is a very large energy compared to SF energy scales, we
introduce a laser detuning $\delta = \omega_0 - \nu_{e g}$.
The operators $\hat{f}_{k \tau}$ are similar to the $f$-modes defined in the
text: $\hat{f}_{k \tau} = \frac{1}{\sqrt{2}} \bigl[ \hat{q}_{k 1} -
\frac{\tau}{r_k} (1 + \eta \e^{\ii k}) \hat{q}_{k 2} \bigr]$.
$\eta$ parameterizes the $e$-atom band structure in the empty tube via
$\epsilon^e_{k \tau} = \tau r_k = \tau \sqrt{1 + \eta^2 + 2 \eta \cos k}$.

The transfer rate ${\cal R}$ can be computed with the aid of the Fermi golden
rule \cite{dao-2009-1}:
\begin{displaymath}
 {\cal R}_\tau (\delta, k) = 2 \pi \sideset{}{_{{\rm in}, \, {\rm fi}}} \sum
 \bigl \vert \langle \psi_{\rm fi} \vert \hat{V}_L \vert \psi_{\rm in} \rangle
 \bigr \vert^2 \,\, \frac{\e^{-E_{\rm in} / T}}{\cal Z} \delta
 ({\cal E}_{\rm fi} - {\cal E}_{\rm in}),
\end{displaymath}
where summation is extended over all initial and final states, and ${\cal Z}$
is the grand partition function.
The matrix elements of $\hat{V}$ are given by
\begin{align}
 \langle \psi_{\rm fi} \vert \hat{V}_L \vert \psi_{\rm in} \rangle = & \Omega
 \sideset{}{_{k a}} \sum \langle \psi^{f, g}_{N_g - 1} \vert \hat{g}_{k a}
 \vert \psi^{i, g}_{N_g} \rangle \hvac \hat{f}_{k \tau} \hat{q}^\dag_{k a} \vac
 = \nonumber \\
 = & {\textstyle \frac{\Omega}{\sqrt{2}}} \bigl[ \langle f \vert \hat{g}_{k 1}
 \vert i \rangle - {\textstyle \frac{\tau}{r_k}} (1 + \eta \e^{\ii k}) \langle
 f \vert \hat{g}_{k 2} \vert i \rangle \bigr], \nonumber
\end{align}
where $\langle f \vert \hat{g}_{k a} \vert i \rangle = \langle \psi^{f, g}_{N_g
- 1} \vert \hat{g}_{k a} \vert \psi^{i, g}_{N_g} \rangle$.

Using the single-particle spectral density \cite{zubarev-1960-1},
\begin{align}
 {\cal A}_{a b} (\omega, k) = & 2 \pi \sideset{}{_{n m}} \sum \langle n \vert
 \hat{g}_{k a} \vert m \rangle \frac{\e^{-E_m / T}}{\cal Z} \nonumber \\
 & \times \langle m \vert \gd_{k b} \vert n \rangle \delta (E_m - E_n -
 \omega), \nonumber
\end{align}
we obtain after a straightforward calculation:
\begin{align}
 {\cal R}_\tau (\delta, k) = & {\textstyle \frac{\Omega^2}{2}} \bigl \lbrace
 {\rm Tr} \, {\cal A} (\omega, k) \nonumber \\
 - {\textstyle \frac{\tau}{r_k}} & \bigl[ (1 + \e^{\ii k}) {\cal A}_{1 2}
 (\omega, k) + (1 + \e^{-\ii k}) {\cal A}_{2 1} (\omega, k) \bigr] \bigr
 \rbrace \nonumber
\end{align}
with $\omega = \epsilon^e_{k \tau} - \mu - \delta$.
Noting that ${\cal A}_{21} = {\cal A}^*_{12}$, we arrive at the expression in
the main text.

\subsection*{Magneto-electric phenomena}

In this section we discuss novel magneto-electric effects that occur as a
result of the {\it odd in momentum} SOC in the effective model (2) when it is
subjected to a weak laser-induced synthetic magnetic field.
We assume this field to be homogeneous: $\delta \hat{H}_{\rm ef} = -{\bm b}
\cdot \sum_k \bsig_{a b} \gd_{k a} \hat{g}_{k b}$.
In particular, we shall derive weak-coupling expressions for the susceptibility
$\chi$ and magneto-electric coefficient $\kappa$ quoted in the main text.

It is well-known that a physical external magnetic field breaks time-reversal
symmetry ${\cal T}$.
However, because ${\bm b}$ is artificial, it does not necessarily break ${\cal
T}$.
Indeed, under time-reversal, the second-quantized operators and $c$-numbers
transform as $\hat{g}_{k a} \to \hat{g}_{-k, a}$ (because $a$ is not a real
spin but the position index inside a dimer) and $c \to c^*$.
Therefore, only the $b_y$-term in $\delta \hat{H}_{\rm ef}$ breaks ${\cal T}$
and is capable of generating a mass current.
Below we focus on the case with $b_x = b_z = 0$ and $b_y \neq 0$ [see Fig.
4(a)].

It is instructive to study the non-interacting system described by the
Hamiltonian $\hat{H}^g_0 + \delta \hat{H}_{\rm ef}$ [$\hat{H}^g_0$ is defined
in Eq. (2)] which is diagonalized by Bogoliubov quasiparticles $\hat{f}_{k
\tau} = \frac{1}{\sqrt{2}} \bigl[ \hat{g}_{k 1} - \frac{\tau}{R_k} (1 +
\e^{-\ii k} - \ii \, b_y) \hat{g}_{k 2} \bigr]$ with energies $\epsilon_{k
\tau} = \tau R_k = \tau \sqrt{2 J_g^2 (1 + \cos k) + b_y^2 + 2 J_g b_y \sin
k}$.
Assuming that $T = 0$ and $f$-particles fill the Fermi sea $\FS$ with a
chemical potential $\mu$, i.e. $\hFS \fd_{k \tau^\prime} \hat{f}_{k \tau} \FS =
\delta_{\tau^\prime \tau} \theta (\mu - \epsilon_{k \tau})$, the average
mass current is (the operator $\hat{K}$ was defined in the main text)
\begin{align}
 \hFS \hat{K} \FS = & {\textstyle \frac{2}{N_d}} \sideset{}{_{k \tau}} \sum
 {\textstyle \frac{\tau}{R_k}} (b_y \cos k - J_g \sin k) \langle \fd_{k \tau}
 \hat{f}_{k \tau} \rangle \nonumber \\
 = {\textstyle \frac{2}{J_g N_d}} & \sideset{}{_{k \tau}} \sum \frac{ \partial
 \epsilon_{k \tau} }{\partial k } \, \theta (\mu \! - \! \epsilon_{k \tau}) =
 {\textstyle \frac{2}{J_g}} (\mu \! - \! \mu) \! = 0. \nonumber
\end{align}
This result is valid even in interacting non-SF systems.
Hence a magnetic field cannot generate a mass current in the absence of SF
correlations \cite{ojanen-2012-1}.

This situation changes dramatically when pairing correlations are taken into
account, technically because the average current can no longer be written as an
integral of the quasiparticle group velocity.
Consider the weak-coupling dilute limit $u_{g g} \ll J_g$ and $n^g \ll 1$.
In this regime, we can project interactions in Eq. (2) onto the lowest band
$\tau = -1$.
At small fields $b_y \ll J_g, \Delta$, the bare fermion operators become
$\hat{g}_{k 1} \approx \frac{1}{\sqrt{2}} \hat{f}_k$ and $\hat{g}_{k 2} \approx
\frac{1}{\sqrt{2}} \e^{\ii k / 2} (1 + \ii b_y / 2 J_g) \hat{f}_k$, where
$\hat{f}_{k, -1} \equiv \hat{f}_k$.
It is easy to check that $b_y$ enters the projected interaction term in Eq. (2)
only via quadratic corrections $\sim b_y^2$.
Therefore, to linear order in $b_y$ the projected Hamiltonian is the same as
Eq. (3) with $\epsilon_k = -2 J_g \cos \frac{k}{2} - b_y \sin \frac{k}{2} =
\epsilon^{(0)}_k - b_y \sin \frac{k}{2}$.
Since $-\epsilon_{-k} = -\epsilon^{(0)}_k - b_y \sin \frac{k}{2}$, the BdG
Hamiltonian is
\begin{displaymath}
 {\cal H}_{\rm BdG} = -b_y \sin {\textstyle \frac{k}{2}} +
 \begin{pmatrix}
  \epsilon^{(0)}_k - \mu & D_k \\
  D_k^* &   \mu-\epsilon^{(0)}_k
 \end{pmatrix}
 ,
\end{displaymath}
see Eq. (S2) in the Supplementary material.
The Bogoliubov transformation diagonalizing this Hamiltonian is given by Eq.
(S3) of the Supplementary material with $\epsilon_k$ and $E_k$ replaced by
$\epsilon^{(0)}_k$ and $E^{(0)}_k = \sqrt{(\epsilon^{(0)}_k - \mu)^2 + \vert
D_k \vert^2}$, respectively.
However, energies of the quasiparticles $\hat{\gamma}_k$ and $\hat{\Gamma}_k$
are now split by a field correction and are given by $E_{k, \pm} = E^{(0)}_k
\mp b_y \sin \frac{k}{2}$, respectively.
If $b_y$ is small, we can assume that ${\rm sign} (E_{k, \pm}) = \pm$, and in
the BCS GS $\langle \hat{\gamma}^\dag_k \hat{\gamma}_k \rangle = \nu (E_{k,
+})$ and $\langle \hat{\Gamma}^\dag_k \hat{\Gamma}_k \rangle = \nu (E_{k, -})$
with the Fermi function $\nu (x) = \bigl( e^{x / T} + 1 \bigr)^{-1}$.

The projected current and $y$-component of the pseudospin $\hat{S}_y$ have the
form: $\hat{K} \approx \frac{1}{N_d} \sum_k \bigl( 2 \sin \frac{k}{2} -
\frac{b_y}{J_g} \cos \frac{k}{2} \bigr) \fd_k \hat{f}_k$ and $\hat{S}_y =
\frac{1}{2 N_d} \sum_k \sigma^y_{a b} \gd_{k a} \hat{g}_{k b} \approx
\frac{1}{4 N_d} \sum_k \bigl( 2 \sin \frac{k}{2} + \frac{b_y}{J_g} \cos
\frac{k}{2} \bigr) \fd_k \hat{f}_k$.
In the BCS GS the first term in both expressions vanishes at $T = 0$ because
$\sideset{}{^\prime} \sum_k \sin \frac{k}{2} \langle \fd_k \hat{f}_k -
\hat{F}^\dag_k \hat{F}_k \rangle = \sideset{}{^\prime} \sum_k \sin \frac{k}{2}
\bigl[ \nu (E_{k, +}) - \nu (E_{k, -}) \bigr]$.
Hence, $\kappa = \langle \hat{K} \rangle / b_y = -4 \langle \hat{S}_y \rangle /
b_y = -4 \chi$.
When $T = 0$, it is easy to show that $\kappa$ is given by the expression in
the main text.

Fig. 4(b) shows $\kappa$ and $\chi$ numerically computed beyond the
weak-coupling dilute limit.
In the simulation we used the same strategy as above: solve the BdG equations
in the field, compute $\langle \hat{K} \rangle$ and $\langle \hat{S}_y \rangle$
as functions of $b_y$, and extract the corresponding linear coefficients.

\section{Supplementary material}
\renewcommand{\theequation}{S\arabic{equation}}

\subsection*{BCS theory in a translationally invariant 1D system}

The unconstrained mean-field approach presented in Methods can be handled only
numerically.
However, if we restrict the variational space to translationally-invariant
states in a chain with periodic boundary conditions (one can use the momentum
$k$-space representation) and omit the Hartree-Fock terms in $\hat{H}_{\rm
HFB}$, the resulting BCS theory can be treated analytically and a number of
insights can be gained on the structure of the topological SF phase.
In particular, in this section we analyze the singlet-triplet mixing of Cooper
pair states due to the SOC in Eq. (2).

At the BCS level \cite{bardeen-1957-1}, the linearized interaction
$\hat{H}_{\rm int}$ in Eq. (M1) contains only the $\hat{h}_B$ term:
\begin{displaymath}
 \hat{H}_{\rm int} \approx \sideset{}{_k} \sum \bigl[ \Delta \, \gd_{k 1}
 \gd_{-k, 2} + {\rm h.c.} \bigr], \nonumber
\end{displaymath}
where $\Delta = -\frac{u_{g g}}{N_d} \sum_k \langle \hat{g}_{-k, 2} \hat{g}_{k
1} \rangle \equiv -\ii \Delta_0 \e^{\ii \phi_\Delta}$ (real $\Delta_0$) is an
$s$-wave OP.
Because the above interaction couples fermions with opposite momenta, to avoid
overcounting of $k$-states we will restrict momentum summations to a
``positive'' half of the BZ with $k > 0$ (indicated by a prime), and denote
$\hat{g}_{-k, a} \equiv \hat{G}_{k a}$ ($-k$ belongs to the other half of the
BZ): $\sum_k f (\hat{g}_{k a}, \hat{g}_{-k, b}) = \sum_k^\prime \bigl[ f
(\hat{g}_{k a}, \hat{G}_{k b}) + f (\hat{G}_{k a}, \hat{g}_{k b}) \bigr]$.
Then
\begin{align}
 \hat{H}_{\rm int} \approx & \sideset{}{^\prime_k} \sum \bigl[ \Delta \, \bigl(
 \gd_{k 1} \hat{G}^\dag_{k 2} + \hat{G}^\dag_{k 1} \hat{g}^\dag_{k 2} \bigr) +
 {\rm h.c.} \bigr] \nonumber \\
 & = \Delta_0 \, \sigma^y_{a b} \sideset{}{^\prime_k} \sum \bigl[
 \e^{\ii \phi_\Delta} \, \gd_{k a} \hat{G}^\dag_{k b} - {\rm h.c.} \bigr].
 \nonumber
\end{align}
Similarly, the kinetic energy in Eq. (2) can be written as
\begin{displaymath}
 \hat{H}^g_0 = \sideset{}{_k} \sum \epsilon_{a b} (k) \gd_{k a} \hat{g}_{k b}
 = \sideset{}{^\prime_k} \sum \epsilon_{a b} (k) \bigl[ \gd_{k a} \hat{g}_{k b}
 - \hat{G}_{k a} \hat{G}^\dag_{k b} \bigr]
\end{displaymath}
because $\epsilon (k) = -J_g \bigl[ (1 + \cos k) \sigma^x + \sin k \, \sigma^y
\bigr] = \epsilon^T (-k)$.

Up to constant terms, the full BCS Hamiltonian is
\begin{displaymath}
 \hat{H}_{\rm BCS} \! = \!\! \sideset{}{^\prime} \sum_k \!
 \bigl( \gd_{k a} \, \hat{G}_{k a} \bigr) \!
 \underbrace{
  \begin{pmatrix}
   \epsilon_{a b} - \mu \delta_{a b} & \Delta_0 \e^{\ii \phi_\Delta}
   \sigma^y_{a b} \\
   \Delta_0 \e^{-\ii \phi_\Delta} \sigma^y_{a b} & \mu \delta_{a b} -
   \epsilon_{a b}
  \end{pmatrix}
 }_{
  {\cal H}_{\rm BdG}
 }
 \!\!
 \begin{pmatrix}
  \hat{g}_{k, b} \\
  \hat{G}^\dag_{k, b}
 \end{pmatrix}
 .
\end{displaymath}
The BdG Hamiltonian ${\cal H}_{\rm BdG}$ can be easily diagonalized when the
OP $\Delta$ is purely imaginary.
This scenario is still quite general, and below we will focus on the case
$\phi_\Delta = 0$ when ${\cal H}_{\rm BdG} = (\sigma^z \otimes (\epsilon-\mu))
+ \Delta_0 (\sigma^x \otimes \sigma^y)$.
It is diagonalized by performing a unitary transformation ${\cal H}_{\rm BdG}
\to {\cal H}^y_{\rm BdG} = (U^\dag \otimes 1) {\cal H}_{\rm BdG} (U \otimes 1)$
with
$U = \frac{1}{\sqrt{2}}
 \begin{pmatrix}
  1 & 1 \\
  \ii & -\ii
 \end{pmatrix}
$.
The eigenstates $\psi_{\nu \tau}$ are labeled by two indices $\tau$ and $\nu =
\pm 1$, and have the form
\begin{equation}
 \psi_{\nu \tau} = \frac{1}{2 T_{k \tau}}
 \begin{pmatrix}
  \bigl[ \nu T_{k \tau} + h_k \bigr] u_{k \tau} \\
  \ii \bigl[ \nu T_{k \tau} - h_k \bigr] u_{k \tau}
 \end{pmatrix}
 \label{eq:BdG_wf_1d}
\end{equation}
with $h_k = \epsilon (k) - \mu + \ii \Delta_0 \sigma^y$ and the spinor
\begin{displaymath}
 u_{k \tau} = \frac{1}{\sqrt{2 R_k (R_k + \tau D_k)}}
 \begin{pmatrix}
  R_k + \tau D_k \\
  \tau \mu (1 + \e^{\ii k})
 \end{pmatrix}.
\end{displaymath}
The corresponding energies are $E_{k \nu \tau} = \nu T_{k \tau}$ with $T_{k
\tau} = \sqrt{\mu^2 + \Delta_0^2 + 2 J_g^2 (1 + \cos k) + 2 \tau J_g R_k}$,
$R_k = \sqrt{D^2_k + 2 \mu^2 (1 + \cos k)}$ and $D_k = \Delta_0 (1 + \cos k)$.

The parameter $\Delta_0$ is determined from the BCS self-consistency condition:
\begin{align}
 \Delta_0 = & -\frac{u_{g g}}{N_d} \sideset{}{^\prime_k} \sum \sigma^y_{a b}
 \langle \hat{g}_{k, a} \hat{G}_{k, b} \rangle
 \nonumber \\
 & = \frac{\Delta_0 u_{g g}}{4 N_d} \sideset{}{_{k \tau}} \sum \frac{{\rm th}
 \, \frac{\beta T_{k \tau}}{2}}{T_{k \tau}} \biggl[ 1 + \tau \frac{J_g}{R_k} (1
 + \cos k)^2 \biggr] \nonumber
\end{align}
with the inverse temperature $\beta = 1 /T$.
To solve this equation, let us assume that the Fermi level (corresponding to
momentum $k_F$) lies below zero, i.e. $\mu = -J_g \sqrt{2 (1 + \cos k_F)} < 0$.
Near the Fermi level $\delta k = \vert k \vert - k_F$, $T_{k, +1} = 2 \vert
\mu \vert$ and $T_{k, -1} = \sqrt{\Delta_0^2 + v_F^2 \delta k^2}$, with Fermi
velocity $v_F = J_g \sqrt{1 - (\mu / 2 J_g)^2}$.
In the weak coupling and zero-temperature limits, the main contribution to the
sum comes from around the Fermi surface:
\begin{align}
 1 \approx & \frac{u_{g g}}{4 N_d} \sum_k \frac{1}{T_{k, -1}} \biggl[ 1 -
 \frac{J_g}{R_k} (1 + \cos k)^2 \biggr]_{k = k_F} \!\! = \frac{u_{g g}}{4}
 \frac{2}{2 \pi} \nonumber \\
 \times & \int_0^{\lesssim k_F} \frac{d k}{\sqrt{\Delta_0^2 + v_F^2 \delta
 k^2}} \frac{v_F^2}{J_g^2} = \frac{u_{g g}}{2 \pi J_g} \sqrt{1 -
 \frac{\mu^2}{(2 J_g)^2}} \ln \frac{2 \rho}{\Delta_0}. \nonumber
\end{align}
From here, $\Delta_0 = 2 \rho \e^{-2 \pi J_g / u_{g g} \sqrt{1 - (\mu / 2
J_g)^2}}$ with $\rho \sim J_g \gg \Delta_0$ -- a characteristic energy measured
from the Fermi level beyond which the above expansion becomes invalid.

The OP $\Delta_0$ involves $g$-fermions with opposite pseudospin $a$ and
describes an $s$-wave Cooper pairing.
However, due to the SOC inherent in $\hat{H}^g_0$, there is also an admixture
of the $p$-wave pairing states of two $g$-atoms with the same pseudospin which,
because of Fermi statistics, must be antisymmetric in momentum.
To quantify these $p$-wave correlations, we compute a spin-averaged pairing
amplitude ${\cal P} = \frac{1}{2 \Delta_0} \sum_a \langle \hat{g}_{k a}
\hat{G}_{k a} \rangle$ using the BdG wavefunctions \eqref{eq:BdG_wf_1d}:
\begin{displaymath}
 {\cal P} \! = \! \frac{\mu \sin k}{4 R_k} \sideset{}{_\tau} \sum
 \frac{\tau}{T_{k \tau}} {\rm th} \, \frac{\beta T_{k \tau}}{2}.
\end{displaymath}
This result demonstrates that $p$-wave correlations in the system are
enslaved to the $s$-wave OP.
Moreover, the $\tau$-dependence of the quasiparticle energy is essential: if
$T_{k, +1} = T_{k, -1}$, as it would be in the absence of SOC, there is no
$p$-wave pairing and ${\cal P} \equiv 0$.

Finally, we derive the BCS approximation in the weak-coupling limit described
by Eq. (3).
Introducing the pairing OP $\Delta = -\frac{u_{g g}}{4 N_d} \sum_k \e^{-\ii
\frac{k}{2}} \langle \hat{f}_{-k} \hat{f}_k \rangle$, one can linearize the
Hamiltonian (3):
\begin{align}
 \hat{H}_{\rm ef} \approx & \sideset{}{_k} \sum \bigl[ (\epsilon_k - \mu) \fd_k
 \hat{f}_k + \Delta \e^{\ii \frac{k}{2}} \fd_k \fd_{-k} + {\rm h. c.} \bigr]
 \label{eq:H_BCS_f} \\
 = & \sideset{}{^\prime_k} \sum \biggl[ (\epsilon_k - \mu) +
 \begin{pmatrix}
  \fd_k & \!\! \hat{F}_k
 \end{pmatrix}
 \begin{pmatrix}
  \epsilon_k - \mu & D_k \\
  D_k^* & \mu - \epsilon_k
 \end{pmatrix}
 \begin{pmatrix}
  \hat{f}_k \\
  \hat{F}^\dag_k
 \end{pmatrix}
 \biggr],
 \nonumber
\end{align}
where $D_k = 2 \ii \Delta \sin \frac{k}{2}$, $\epsilon_{-k} = \epsilon_k$, as
before $\sum^\prime_k \ldots$ extends over half of the BZ, and $\hat{F}_k =
\hat{f}_{-k}$.
This Hamiltonian is diagonalized by a Bogoliubov transformation from
$f$-fermions to quasiparticles $\hat{\gamma}_k$ and $\hat{\Gamma}_k$
\cite{landau-vol-9}:
\begin{align}
 \begin{pmatrix}
  \hat{f}_k \\
  \hat{F}^\dag_k
 \end{pmatrix}
 = & \frac{1}{\sqrt{2 E_k}} \biggl[ \frac{1}{\sqrt{E_k + (\epsilon_k - \mu)}}
 \begin{pmatrix}
  E_k + (\epsilon_k - \mu) \\
  D^*_k
 \end{pmatrix}
 \hat{\gamma}_k \nonumber \\
 - & \frac{1}{\sqrt{E_k - (\epsilon_k - \mu)}}
 \begin{pmatrix}
  E_k - (\epsilon_k - \mu) \\
  -D^*_k
 \end{pmatrix}
 \hat{\Gamma}^\dag_k
 \biggr].
\label{eq:H_BCS_f_qp}
\end{align}
Here $E_k = \sqrt{(\epsilon_k - \mu)^2 + \vert D_k \vert^2}$ is the
quasiparticle dispersion.
The SF gap has a $p$-wave symmetry: $D_k = -D_{-k}$ and its amplitude $\Delta$
is determined from the self-consistency equation $1 = \frac{u_{g g}}{4 N_d}
\sum_k \sin^2 \frac{k}{2} / E_k$: $\Delta \sim \e^{-2 \pi J_g / u_{g g} \sqrt{1
- (\mu / 2 J_g)^2}}$, which coincides with the result obtained in the
multi-band case.

\subsection*{Topological superfluidity in a 2D lattice}

\begin{figure}[t]
 \begin{center}
  \includegraphics[width = \columnwidth]{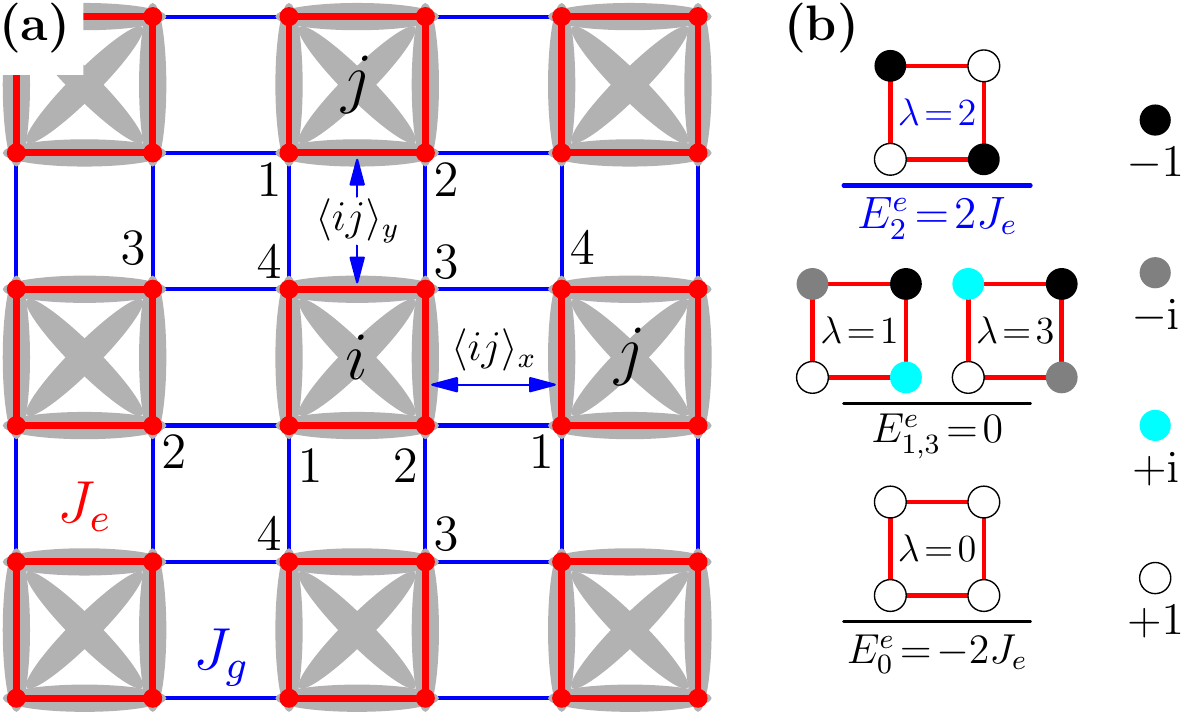}
 \end{center}
 \caption{
  {\bf (a)} Two-dimensional optical superlattice with a plaquette structure.
  The $e$-atoms are confined within red square plaquettes (with one atom per
  cell), $g$-atoms propagate in a simple square lattice (thin blue lines).
  Gray ellipses denote $g$-atom attraction mediated by fluctuations of the
  $e$-subsystem.
  Other notations are as in Fig. 1(a).
  {\bf (b)} States of an $e$-atom within a plaquette.
  Also shown are point symmetries of the wavefunctions $X^{(\lambda)}_a =
  \frac{1}{2} \e^{\ii \frac{\pi}{2} \lambda (a - 1)}$ with colors indicating
  phases $\pm 1$ and $\pm \ii$.
  In particular, the $\lambda = 0$ ($2$) state has $s$- ($d$-) wave symmetry.
  Attractive interactions between $g$-atoms occur when $e$-atoms are prepared
  in the $\lambda = 2$ state.
 }
 \label{fig:sfig1}
\end{figure}

The results presented in the main text, can be extended to higher dimensional
systems.
Here we briefly consider a generalization involving a lattice that has the
plaquette structure shown in Fig. \ref{fig:sfig1}(a).
In each plaquette, there is one $e$-atom that can tunnel around the plaquette
(with lattice constant $a_0$).
On the other hand, $g$-fermions move in the simpler square lattice.
As before, all atoms are nuclear-spin-polarized and we use units where $a_0 =
1$.
The Hamiltonian of this system is
\begin{align}
 \hat{H} = & \hat{H}^e_0 + \hat{H}^g_0 + \hat{H}_{\rm int};
 \label{eq:H_2d} \\ 
 \hat{H}^e_0 = & -J_e \sum_{i, \, \Box_{a b}} \bigl( \ed_{i a} \hat{e}_{i b} +
 {\rm h. c.} \bigr); \nonumber \\
 \hat{H}^g_0 = & -J_g \biggl[ \,\, \sum_{i, \, \Box_{a b}} \bigl( \gd_{i a}
 \hat{g}_{i b} + {\rm h.c.} \bigr) \nonumber \\
 + & \!\! \sum_{\langle i j \rangle_x} \! \bigl( \gd_{i 2} \hat{g}_{j 1} +
 \gd_{i 3} \hat{g}_{j 4} \bigr) \! + \!\! \sum_{\langle i j \rangle_y} \!
 \bigl( \gd_{i, 4} \hat{g}_{j, 1} + \gd_{i 3} \hat{g}_{j 2} \bigr) \! +
 \! {\rm h. c.} \biggr]; \nonumber \\
 \hat{H}_{\rm int} = & \,\, U^-_{e g} \sum_{i a} \hat{n}^g_{i a}
 \hat{n}^e_{i a}, \nonumber
\end{align}
where $i = \bx_i = (x_i, y_i)$, $i$, $j = 1, \ldots, N_\Box$ label plaquettes
in the lattice, $a$, $b = 1, 2, 3, 4$ denote wells inside a plaquette,
$\Box_{a b}$ indicates all sides $\langle a b \rangle$ of a square, and
$\langle i j \rangle_{x, y}$ is a link connecting two plaquettes in the $x$ or
$y$ direction [Fig. \ref{fig:sfig1}(a)].
Other notations are the same as in Eq. (1).

In the rest of this section, we shall follow an analysis similar to the one in
the main text:
First, we consider $e$-atom states of an isolated plaquette and identify a
spatial mode that gives rise to the $g$-atom attraction within that plaquette.
Then, we study a weak-coupling dilute limit $u_{g g} \ll J_g$, $n^g \ll 1$ and
demonstrate the stability of a chiral $p_x + \ii p_y$ TSF state.
The study of the phase diagram, magneto-electric effects, etc. is left for a
future investigation.

\paragraph{Pairing of the $g$-atoms.--}
An $e$-atom localized within a plaquette, has four states $\lambda = 0, \ldots,
3$ shown in Fig. \ref{fig:sfig1}(b).
Their wavefunctions are $\vert \lambda \rangle_i = \sum_a X^{(\lambda)}_a
\ed_{i a} \vac$ with $X^{(\lambda)}_a = \frac{1}{2} \e^{\ii \frac{\pi}{2}
\lambda (a - 1)}$ and corresponding energies $E^e_0 = -E^e_2 = -2 J_e$, $E^e_1
= E^e_3 = 0$.
Matrix elements of the density $\hat{n}^e_{i a}$ are ${}_i \langle
\lambda^\prime \vert \hat{n}^e_{i a} \vert \lambda \rangle_i =
X^{(\lambda^\prime) *}_a X^{(\lambda)}_a = \frac{1}{4} \e^{\ii \frac{\pi}{2}
(\lambda - \lambda^\prime) (a - 1)}$.

In the 1D case we saw that an attractive interaction between $g$-atoms occurs
when the $e$-subsystem is prepared in the highest excited kinetic energy state.
An analogous situation happens here: $e$-atoms must fluctuate out of the
$d$-wave $\lambda = 2$ state with energy $2 J_e$.
Using the Schrieffer-Wolff transformation derived in Methods and performing
similar calculations, we obtain the second-order $g$-atom Hamiltonian
\begin{align}
 \hat{H}_{\rm ef} = & \hat{H}^g_0 - u_{g g} \sideset{}{_i} \sum \bigl[
 (\hat{n}^g_{i 1} + \hat{n}^g_{i 3}) (\hat{n}^g_{i 2} + \hat{n}^g_{i 4})
 \nonumber \\
 & + 3 (\hat{n}^g_{i 1} \hat{n}^g_{i 3} + \hat{n}^g_{i 2} \hat{n}^g_{i 4})
 \bigr].
 \label{eq:H_ef-2d}
\end{align}
Here $u_{g g} = (U^-_{e g})^2 / 32 J_e$, and density terms in the 1st (2nd)
line describe attractive interactions along sides (diagonals) of the $i$-th
plaquette [gray ellipses in Fig. \ref{fig:sfig1}(a)].

\paragraph{Weak-coupling dilute regime.--}
The $g$-atom kinetic energy can be written as
\begin{displaymath}
 \hat{H}^g_0 = -J_g \sum_\bk \gd_{\bk a}
 \begin{pmatrix}
  0 & \eta_{k_x} & 0 & \eta_{k_y} \\
  \eta^*_{k_x} & 0 & \eta_{k_y} & 0 \\
  0 & \eta^*_{k_y} & 0 & \eta^*_{k_x} \\
  \eta^*_{k_y} & 0 & \eta_{k_x} & 0
 \end{pmatrix}
 _{\!\!\!\! a b} \hat{g}_{\bk b},
\end{displaymath}
where $\hat{g}_{\bk a} = \frac{1}{\sqrt{N_\Box}} \sum_i \e^{-\ii (\bk \cdot
\bx_i)} \hat{g}_{i a}$, momentum $\bk = (k_x, k_y)$ with $k_\alpha \in [-\pi,
\pi]$ ($\alpha = x$, $y$) is defined in a plaquette BZ with $N_\Box$ states and
$\eta_{k_\alpha} = 1 + \e^{-\ii k_\alpha}$.
Eigenvalues of the matrix in $\hat{H}_0^g$ are $\pm 2 J_g \bigl[ \cos
\frac{k_x}{2} \pm \cos \frac{k_y}{2} \bigr]$.
At weak coupling $u_{g g} \ll J_g$ and $n^g \ll 1$, only the lowest band
$\epsilon_\bk = - 2 J_g \bigl[ \cos \frac{k_x}{2} + \cos \frac{k_y}{2} \bigr]
\approx -4 J_g + \frac{k^2}{2 m^*}$ is populated ($m^* = 2 / J_g$ is the
effective mass).
The corresponding eigenvector is $\psi_0 = \frac{1}{2} \bigl( 1, \,
-\e^{-\ii \varphi_x}, \, \e^{-\ii (\varphi_x + \varphi_y)}, \,
-\e^{-\ii \varphi_y} \bigr)^T$ with $\e^{\ii \varphi_\alpha} = \eta_{k_\alpha}
/ \vert \eta_{k_\alpha} \vert$.
At small momenta, $\psi_0 \approx \frac{1}{2} \bigl( 1, \, \e^{\ii k_x / 2}, \,
\e^{\ii (k_x + k_y) / 2}, \, \e^{\ii k_y / 2} \bigr)^T$.

The low-energy effective Hamiltonian can be obtained by replacing $\hat{g}_\bk
\to \psi_0 f_\bk$ ($f_\bk$ is the lowest-band fermion quasiparticle) in Eq.
\eqref{eq:H_ef-2d}:
\begin{displaymath}
 \hat{H}_{\rm ef} \approx \sideset{}{_\bk} \sum \epsilon_\bk \fd_\bk
 \hat{f}_\bk + \frac{u_{g g}}{16 N_\Box} \sideset{}{_{\bk^\prime \bk \bq}} \sum
 \bq^2 \fd_{\bk + \bq} \fd_{\bk^\prime - \bq} \hat{f}_{\bk^\prime} \hat{f}_\bk.
\end{displaymath}
This expression describes a system of spinless fermions interacting via an
attractive $p$-wave coupling.
Within the BCS approximation we have:
\begin{align}
 \hat{H}_{\rm ef} \approx & \sideset{}{_\bk} \sum \bigl[ \xi_\bk \fd_\bk
 \hat{f}_\bk + {\textstyle \frac{1}{2}} (\bk \cdot {\bm \Delta}) \fd_\bk
 \fd_{-\bk} + {\rm h. c.} \bigr] \nonumber \\
 & = \sideset{}{^\prime_\bk} \sum \biggl[ \xi_\bk +
 \begin{pmatrix}
  \fd_\bk & \hat{F}_\bk
 \end{pmatrix}
 \begin{pmatrix}
  \xi_\bk & D_\bk \\
  D^*_\bk & -\xi_\bk
 \end{pmatrix}
 \begin{pmatrix}
  \hat{f}_\bk \\
  \hat{F}^\dag_\bk
 \end{pmatrix}
 \biggr]. \nonumber
\end{align}
Here we switched to a grand-canonical ensemble with a chemical potential $\mu$
and performed the standard replacement $\hat{H}_{\rm ef} \to \hat{H}_{\rm ef} -
\mu \sum_{i a} \hat{n}^g_{i a}$.
$\xi_\bk = \epsilon_\bk - \mu$, $D_\bk = (\bk \cdot {\bm \Delta})$, and other
notations are the same as in the 1D weak-coupling case [see Eq. (1) in the main
text and Methods].
The above Hamiltonian is diagonalized by a Bogoliubov transformation identical
to the one used in the 1D case after a replacement $k \to \bk$.
The pairing OP ${\bm \Delta} = -\frac{u_{g g}}{2 N_\Box} \sum^\prime_\bk \bk
\langle \hat{F}_\bk \hat{f}_\bk \rangle$ obeys the BCS equation
\begin{equation}
 {\bm \Delta} = \frac{u_{g g}}{4 N_\Box} \sideset{}{^\prime_\bk} \sum
 \frac{\bk (\bk \cdot {\bm \Delta})}{E_\bk} = \frac{u_{g g}}{8 N_\Box}
 \sideset{}{_\bk} \sum \frac{\bk (\bk \cdot {\bm \Delta})}{E_\bk},
 \label{eq:bcs-2d}
\end{equation}
where $E_\bk = \sqrt{\xi_\bk^2 + \vert D_\bk \vert^2}$ and in the last sum we
used evenness of the integrand and extended summation over the entire BZ.
Whenever the SF phase develops, the grand potential ${\cal G}_s$ in that phase
is reduced compared to its normal-state value ${\cal G}_n$.
This shift can be computed using the Hellman-Feynman theorem
\cite{landau-vol-9}
\begin{displaymath}
 \frac{\delta {\cal G}}{N_\Box} = \frac{{\cal G}_s - {\cal G}_n}{N_\Box} =
 \int_0^{u_{g g}} \! d \widetilde{u} \, \frac{\langle \hat{H}_{\rm int} \rangle}
 {N_\Box \, \widetilde{u}} = -2 \int_0^{u_{g g}} \! d \widetilde{u} \,
 \frac{\vert \Delta (\widetilde{u}) \vert^2}{\widetilde{u}^2}.
\end{displaymath}
There are two competing SF states characterized by different symmetries
of the OP ${\bm \Delta}$: (i) chiral $p_x \pm \ii p_y$ phase with ${\bm
\Delta}_{p + \ii p} = \Delta_1 ({\bm e}^p_x \mp \ii {\bm e}^p_y)$ ($\bp = p_x
{\bm e}^p_x + p_y {\bm e}^p_y$), and (ii) $p_x$ (or $p_y$) state with ${\bm
\Delta}_{p_x} = \Delta_2 {\bm e}^p_x$ (or $\Delta_2 {\bm e}^p_y$).
Below we assume real amplitudes $\Delta_{1, 2}$ and demonstrate that at weak
coupling, the $p_x + \ii p_y$ state is favored.

\begin{figure}[t]
 \begin{center}
  \includegraphics[width = \columnwidth]{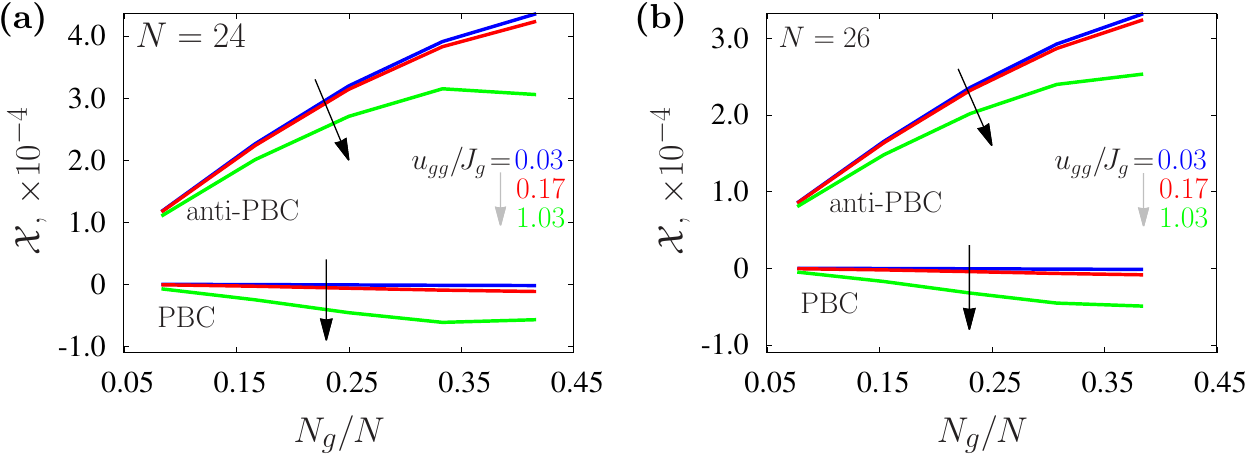}
 \end{center}
 \caption{
  Inverse compressibility ${\cal X}$ of a chain with periodic and anti-periodic
  boundary conditions (PBC and anti-PBC) computed using exact diagonalization
  for {\bf (a)} $N = 24$ and {\bf (b)} $N = 26$ sites.
  The number of $g$-atoms is even: $N_g = 2, \ldots, 10$.
  Red, blue and green curves correspond to interaction strengths $u_{g g} / J_g
  = 0.03$, $0.17$ and $1.03$, respectively.
  Arrows show direction of increasing $u_{g g}$ in the above interval.
 }
 \label{fig:sfig2}
\end{figure}

\paragraph{$p_x + \ii p_y$ SF.--}
The $u_{g g}$-dependence of $\Delta_1$ can be determined by multiplying Eq.
\eqref{eq:bcs-2d} by ${\bm \Delta}^*$ [notice that $\vert {\bm \Delta} \vert^2
= 2 \Delta_1^2$ and $\vert \bk \cdot {\bm \Delta} \vert^2 = \bk^2 \Delta_1^2$]:
$2 = \frac{u_{g g}}{8 N_\Box} \sum_\bk \frac{\bk^2}{E_\bk} = \frac{u_{g g}}{16
\pi} \int d k \frac{k^3}{E_\bk} \approx \kappa \ln \frac{2 \rho}{k_F
\Delta_1}$, where $\kappa = u_{g g} Q_F$ with $Q_F \equiv \frac{m^* k_F^2}{8
\pi}$, $k_F$ is the Fermi momentum, and $\rho \sim J_g \gg \Delta_1$ is a
characteristic energy cutoff [cf. Methods]. We have:
\begin{displaymath}
 \frac{\delta {\cal G}_1}{N_\Box} = 4 Q_F \int_0^{\Delta_1} d \bigl(
 {\textstyle \frac{1}{\kappa}} \bigr) \Delta^2 = -Q_F \Delta_1^2 =
 -\frac{\rho^2 m^*}{2 \pi} \e^{-4 / \kappa}.
\end{displaymath}

\paragraph{$p_x$ SF.--}
Proceeding in a similar manner as above, we use the BCS equation
\eqref{eq:bcs-2d} to calculate $\Delta_2$
\begin{align}
 1 = & \frac{u_{g g}}{8 N_\Box} \! \sum_\bk \frac{k_x^2}{E_\bk} =
 \frac{u_{g g}}{8} \int \frac{d k \, k}{2 \pi} \oint \frac{d \theta}{2 \pi}
 \frac{k^2 \cos^2 \theta}{\sqrt{\xi_\bk^2 + k^2 \Delta_2^2 \cos^2 \theta}}
 \nonumber \\
 \approx & \kappa \oint \frac{d \theta}{2 \pi} \, \cos^2 \theta \, \ln
 \frac{2 \rho}{k_F \Delta_2 \vert \cos \theta \vert} = \frac{\kappa}{2} \biggl[
 \ln \frac{2 \rho}{k_F \Delta_2} + I \biggr], \nonumber
\end{align}
where $I = -2 \oint \frac{d \theta}{2 \pi} \cos^2 \theta \ln \vert \cos \theta
\vert \approx 0.2$.
$\delta {\cal G}$ can be computed as before
\begin{align}
 \frac{\delta {\cal G}_2}{N_\Box} = & 2 Q_F \int_0^{\Delta_2} d \bigl(
 {\textstyle \frac{1}{\kappa}} \bigr) \Delta^2 = -\frac{Q_F \Delta_2^2}{2} =
 -\frac{\rho^2 m^*}{2 \pi} \frac{\e^{2 I}}{2} \e^{-4 / \kappa}.
 \nonumber
\end{align}
Because $\e^{2 I} / 2 < 1$, $\delta {\cal G}_1 < \delta {\cal G}_2$ and the
$p_x \pm \ii p_y$ SF state is preferred over the striped $p_{x, y}$ phase.

\subsection*{Fermion parity switch in the Hamiltonian (2)}

In this section, we show that the exact GS of the effective Hamiltonian (2)
realizes a fermionic parity switch \cite{ortiz-2014-1,ortiz-2016-1}, i.e. the
inverse compressibility, ${\cal X} = \frac{1}{2 N} \bigl[ E_0 (N_g + 1) + E_0
(N_g - 1) \bigr] - E_0 (N_g)$, changes sign when the boundary conditions are
switched from periodic to anti-periodic, thus confirming the topological nature
of our system beyond mean-field.
In Fig. \ref{fig:sfig2} we present results of the exact diagonalization (using
the Lanczos technique of Ref. \cite{wu-2000-1}) in systems with $N = 24$ and
$26$ sites (12 and 13 dimers, respectively) and even number of $g$-atoms $N_g =
2, \ldots, 10$.
Clearly, ${\cal X}$ has different sign for the two types of boundary
conditions.